\documentclass[a4paper,11pt]{article}
\usepackage{cite}
\usepackage[all]{xy}
\usepackage{amsmath,amsfonts,amssymb,amsthm,epsfig,amscd,comment,latexsym,psfrag}
\usepackage{CJK}
\usepackage{mathrsfs}
\usepackage{nicefrac,xspace,tikz}
\usepackage{arydshln}
\usepackage{stmaryrd}
\usepackage{graphicx,bm}
\usepackage{amscd}
\usetikzlibrary{arrows}

\makeatletter

\newcommand{\Rmnum}[1]{\expandafter\@slowromancap\romannumeral #1@}

\makeatother
\usepackage{graphicx,amssymb,amsmath,bm,latexsym}
\allowdisplaybreaks

\newtheorem{theorem}{Theorem}
\numberwithin{equation}{section}

\begin{document}{
\begin{titlepage}
\begin{center}
{\Large\bf Superintegrability for some $(q,t)$-deformed matrix models}\vskip .2in
{\large Fan Liu$^{a,}$\footnote{liufan-math@cnu.edu.cn},
		Rui Wang$^{b,}$\footnote{wangrui@cumtb.edu.cn},
		Jie Yang$^{a,}$\footnote{yangjie@cnu.edu.cn}
		and Wei-Zhong Zhao$^{a,}$\footnote{Corresponding author: zhaowz@cnu.edu.cn}} \vskip .2in
		$^a${\em School of Mathematical Sciences, Capital Normal University, Beijing 100048, China} \\
		$^b${\em Department of Mathematics, China University of Mining and Technology, Beijing 100083, China}\\
\begin{abstract}
We analyze the Macdonald's $(q,t)$-deformed hypergeometric functions with one and two set variables and present their constraints.
We prove the uniqueness to the solutions of these constraints. We propose a concise method to prove the superintegrability relations for
$(q,t)$-deformed matrix models, where the constraints of hypergeometric functions play a crucial role.
A conjectured superintegrability relation in the literature for the refined Chern-Simons model can be easily proved by our method.
Moreover, we construct a general $(q,t)$-deformed matrix model. We give the constraint conditions for parameters in the integral.
The superintegrability relations for the $(q,t)$-deformed integrals with allowed parameters are derived from the hypergeometric constraints.
\end{abstract}
\end{center}		
\end{titlepage}

\section{Introduction}

Matrix models provide a rich set of approaches to physical systems. Already a considerable amount is known about $W$-representations of matrix models,
which realize the partition functions by acting on elementary functions with exponents of the given $\hat{W}$-operators \cite{Morozov09,Cassia,MirGKM,Ale10,Ale18}.
Much investigations have been made for the superintegrability of matrix models \cite{MirSum} by means of $W$-representations. Here the superintegrability means that
for the matrix models, the average of a properly chosen symmetric function is proportional to ratios of symmetric functions on a proper locus,
i.e., $\langle character\rangle \sim character$.

Hypergeometric Hurwitz $\tau$-functions associated with the Hurwitz counting on the Riemann surface are closely related to the matrix models \cite{Alexandrov14,Alexandrovchain,H1,H2,H3}.
These $\tau$-functions belong to KP/Toda integrable hierarchy \cite{Orlov1,Orlov2} and can be described by certain matrix models with $W$-representations and  superintegrability
\cite{Alexandrov14}. A family of $\beta$-deformed (skew) hypergeometric Hurwitz $\tau$-functions have been constructed by $W$-representations. Their integral realizations
and superintegrability relations have been studied \cite{Rui,MirInter,MirSkew,MirCom,Oreshina1,Oreshina2}. By the generalized Laplace transformation of Jack polynomials \cite{Baker97},
some $\beta$-deformed multi-matrix integrals with superintegrability were constructed \cite{Fan}, which belong to the family of hypergeometric functions \cite{MacdonaldI}.
The constraints of the $\beta$-deformed hypergeometric functions have been studied \cite{Fan,Chen}.

It would also be possible to lift the process to the $(q,t)$-deformed case \cite{Mor18}. A family of $(q,t)$-deformed (skew) hypergeometric $\tau$-functions can be constructed
by $W$-representations \cite{Fanqt}, where the $W$-operators are given by Ding-Iohara-Miki (DIM) algebra \cite{DI,Miki}. However, it still remains unclear for the integral forms
of these $(q,t)$-deformed partition functions. It is natural to consider the relations between some well-known $(q,t)$-deformed matrix models with superintegrability
and $(q,t)$-deformed hypergeometric functions \cite{MacII}.

There has been the progress in superintegrability for some $(q,t)$-deformed matrix models. The Selberg integral \cite{Selberg}, as the generalization of the Euler beta function,
was initially used to prove some outstanding conjectures in random matrix theory \cite{Forrester08}, then was widely used in orthogonal polynomial theory \cite{AmotoJ}
and conformal field theory \cite{Fateev}. The integral has been evaluated in various forms \cite{Askey,Aomoto,Diejen,Warnaar1,Warnaar2}. The $q$-Selberg integral \cite{Askey,Aomoto},
as the $q$-analogue of Selberg integral, is closely related to the $(q,t)$-deformed hypergeometric functions \cite{MacII,Kaneko2}. The superintegrability relations for the $q$-Selberg
integral were analyzed in Refs. \cite{Kadell,Macdonald,Kaneko2}.

Some $(q,t)$-deformed matrix models associated with the $\mathcal{N}=2$ supersymmetric gauge theory on the 3-manifold $D^2\times_qS^1$ were considered in Ref.\cite{Cassia}.
Some of them can be regarded as the $(q,t)$-analogue of the Laguerre and Gaussian (or Hermite) ensembles. Their superintegrability relations were conjectured by solving the
$q$-Virasoro constraints \cite{Cassia,Nedelin}. Recently, the superintegrability of the $(q,t)$-deformed Gaussian ensemble has been proved \cite{Forrester25} through the theory of
$q$-orthogonal polynomials \cite{Baker00}.

The refined Chern-Simons theory \cite{Aganagic} has been introduced to give a new physical interpretation for certain refined knot invariants which is previously constructed
by homological methods  \cite{Dunfield}. The unknot partition function for refined Chern-Simons model is given by a $(q,t)$-deformed matrix integral \cite{Aganagic2}.
The conjecture for its superintegrability relation was made in Ref.\cite{Cassia22} and still remains open.

By using hypergeometric constraints, Kaneko \cite{Kaneko2,Kaneko1} established the relation between the ($(q,t)$-deformed) Gauss hypergeometric function and the ($q$-deformed) Selberg integral.
This relation implies the superintegrability for the ($q$-deformed) Selberg integral. In this paper, we intend to develop a universal and concise method to prove superintegrability
for other $(q,t)$-deformed matrix models, especially, for the unknot refined Chern-Simons model. First of all, we investigate Macdonald's $(q,t)$-deformed hypergeometric functions.
We construct their constraints by applying the Pieri formulas and certain commutative subalgebra of DIM algebra. Then we consider some $(q,t)$-deformed matrix integrals
and present their constraints by the $q$-analogue of the Stokes' formula formula. Finally, by the uniqueness of the solution to these constraints, we conclude that the averages of
hypergeometric function for these integrals are still hypergeometric functions. These results imply the superintegrability for these $(q,t)$-deformed matrix models.
It should be emphasized that in our concise method, it is not necessary to consider the specific forms of higher order constraint operators for these $(q,t)$-deformed matrix integrals.

This paper is organized as follows. In section 2, we recall the Macdonald polynomials and $(q,t)$-deformed hypergeometric functions.
Then we present constraints of the $(q,t)$-deformed hypergeometric functions and prove uniqueness to the solutions of these constraints.
In section 3, we consider four $(q,t)$-deformed matrix models and prove their superintegrability relations.
In section 4, we propose a general $(q,t)$-deformed matrix model. Under a set of constraint conditions for the parameters in the integral, we obtain four $(q,t)$-deformed models
and analyze their superintegrability. We end this paper with conclusions in section 5.

At the moment of finalizing this paper we became aware that a similar consideration has just appeared in a wonderful paper \cite{CassiaNew}.

\section{($q,t$)-deformed hypergeometric functions and their constraints}

\subsection{Preliminaries associated with Macdonald polynomials}

Let us begin with the Macdonald's difference operators \cite{Macdonald}
\begin{align}\label{DAT}
\mathcal{D}_N^k(\mathbf{x})=\sum_{I\subseteq\{1,\cdots,N\},|I|=k}t^{k(k-1)/2}
\prod_{i\in I,j\notin I}\frac{tx_{i}-x_{j}} {x_{i}-x_{j}}\prod_{i\in I}T_{q,i}(\mathbf{x}),
\end{align}
where $1\le k\le N$, $N\in\mathbb{Z}_+$, $\mathbf{x}=(x_1,\dots,x_N)$ and
$T_{q,i}(\mathbf{x})=q^{x_i\frac{\partial}{\partial x_i}}$.

Moreover, the generating function of (\ref{DAT}) is given by
\begin{align}
	\mathcal{D}_N(z;\mathbf{x})=\frac{\det_{1\le i,j\le N} \left(x_i^{N-j}(1+zt^{N-j}T_{q,i})\right)}{\det_{1\le i,j\le N} \left(x_i^{N-j}\right)}=\sum_{k=1}^N\mathcal{D}^{k}_N(\mathbf{x})z^{k},
\end{align}
which satisfies
\begin{eqnarray}\label{Dqt}
	\mathcal{D}_N(z;\mathbf{x})J^{(q,t)}_{\lambda}(\mathbf{x})=\prod_{i=1}^N (1+zq^{\lambda_i}t^{N-i})J^{(q,t)}_{\lambda}(\mathbf{x}),
\end{eqnarray}
where $\lambda=(\lambda_1,\lambda_2,\cdots)$ are partitions and $J_{\lambda}^{(q,t)}(\mathbf{x})$ are the integral forms of the Macdonald polynomial \cite{Macdonald}.

Let us introduce the Lassalle's operators \cite{Lassalle}
\begin{eqnarray}\label{Edef}
	\mathcal{E}_k(\mathbf{x})=\sum_{i=1}^{N}x_i^kA_{t,i}(\mathbf{x})\frac{\partial}{\partial_q x_i},
\end{eqnarray}
where $ k\in\mathbb{N}$, $A_{t,i}(\mathbf{x})=\prod_{i\neq j}\frac{tx_i-x_j}{x_i-x_j}$ and $\frac{\partial}{\partial_q x_i}=x_i^{-1}\frac{1-T_{q,i}(\mathbf{x})}{1-q}$
is the $q$-derivative \cite{Exton}.

There are two useful formulas \cite{Awata}
\begin{subequations}\label{twouse}
\begin{align}
	\prod_{i=1}^{N}\frac{t-x_iz}{1-x_iz}=&1+\sum_{i=1}^{N}\frac{t-1}{1-x_iz}A_{t,i}(\mathbf{x}),\label{ATA}\\
	T_{q,i}(\mathbf{x})p_n(\mathbf{x})=&(q^n-1)x_i^n+ p_n(\mathbf{x})T_{q,i}(\mathbf{x}),
\end{align}
\end{subequations}
where $p_n(\mathbf{x})=\sum_{i=1}^Nx_i^n$ for $n\in\mathbb{Z}$ are the power sum functions. The formula (\ref{ATA}) can be checked by the residue theorem
at the singular points $z=\infty$ and $z=x_i^{-1}$ for $i=1,\cdots,N$.

It follows from the formulas (\ref{twouse}) that
\begin{subequations}\label{twouse2}
\begin{align}
	\sum_{i=1}^{N}\frac{A_{t,i}(\mathbf{x})}{1-x_iz}=&\frac{1}{t-1}\left[t^N\exp\left(\sum_{n=1}^{\infty}\frac{1-t^{-n}}{n}p_n(\mathbf{x})z^n\right)-1\right],\label{Axk}\\
	x_i^kT_{q,i}(\mathbf{x})=&\oint\frac{dz}{2\pi \mathbf{i}z} \frac{z^{-k}}{1-x_iz} \exp\left(-\sum_{n=1}^{\infty} (1-q^n)\frac{\partial}{\partial p_n(\mathbf{x})}z^{-n}\right).
\end{align}
\end{subequations}
Therefore, the Lassalle's operators
$\mathcal{E}_k(\mathbf{x})$ (\ref{Edef}) can be rewritten by the collective variables $\mathbf{p}=(p_1,p_2,\cdots)$
\begin{align}\label{Ecoll}
	\mathcal{E}_k\{\mathbf{p}\}=&\frac{1}{(1-q)(t-1)}\oint\frac{dz}{2\pi \mathbf{i}}z^{-k}\left[t^N\exp\left( \sum_{n=1}^{\infty}\frac{1-t^{-n}}{n}p_nz^n\right)-1 \right]\nonumber\\
	&\times\left[1-\exp\left( -\sum_{n=1}^{\infty} (1-q^n)\frac{\partial}{\partial p_n} z^{-n}\right)\right],
\end{align}
where we have taken $p_n=\sum_{i=1}^Nx_i^n$. Note that $\mathcal{E}_1\{\mathbf{p}\}$ was first given in Ref.\cite{Awata}.

We denote
\begin{subequations}
\begin{align}
	\mathcal{A}_{\pm k}(\mathbf{x})
	&=\frac{1}{1-q}\sum_{i=1}^N A_{t,i}(\mathbf{x})x_i^{\pm k},\label{Akxx}\\
	\bar{\mathcal{A}}_{\pm k}(\mathbf{x})&=\frac{1}{1-q}\sum_{i=1}^N A_{t^{-1},i}(\mathbf{x})x_i^{\pm k} =\frac{t^{1-N}}{1-q}\sum_{i=1}^N A_{t,i}(\mathbf{x}^{-1})x_i^{\pm k},
\end{align}
\end{subequations}
where $k\in\mathbb{N}$. It follows from (\ref{Axk}) that
\begin{align}\label{Ak+x}
\mathcal{A}_k(\mathbf{x})&=\frac{1}{(1-q)(t-1)}\oint\frac{dz}{2\pi\mathbf{i}z^{k+1}}\left[t^N\exp\left(\sum_{n=1} ^{\infty}
\frac{1-t^{-n}}{n}p_n(\mathbf{x})z^n\right)-1 \right],\nonumber\\
\mathcal{A}_{-k}(\mathbf{x})&=t^{N-1}\bar{\mathcal{A}}_{k}(\mathbf{x}^{-1})=t^{N-1}\left(\mathcal{A}_{k}(\mathbf{x}^{-1})\Big|_{t\to t^{-1}}\right), \quad k\in\mathbb{N}.
\end{align}
We write down (\ref{Akxx}) for $k=0$ and $1$
\begin{align}
	&\mathcal{A}_0(\mathbf{x})=\frac{\{N\}_t}{1-q}, \quad
	\mathcal{A}_1(\mathbf{x}) =\frac{t^{N-1}}{1-q}p_1(\mathbf{x}),\quad \mathcal{A}_{-1}(\mathbf{x}) =\frac{1}{1-q}p_1(\mathbf{x}^{-1}),
\end{align}
where $\{N\}_t=\frac{1-t^N}{1-t}$.

The Macdonald polynomials  $J^{(q,t)}_{\lambda}(\mathbf{x})$ can be extended to the functions with variables $\mathbf{p}$ by
\begin{eqnarray}
	t^{1-N}\mathcal{E}_1\{\mathbf{p}\}J^{(q,t)}_{\lambda}\{\mathbf{p}\}=\left(\sum_{(i,j)\in\lambda}q^{j-1}t^{1-i}\right)J^{(q,t)}_{\lambda}\{\mathbf{p}\}.
\end{eqnarray}
Taking $p_n=\sum_{i=1}^Nx_i^n$ for $n\in\mathbb{N}$, we have
\begin{align}\label{ekpnx}
	\mathcal{E}_k\left\{p_n=\sum_{i=1}^Nx_i^n\right\}=\mathcal{E}_k(\mathbf{x}),\quad
	J^{(q,t)}_{\lambda}\left\{p_n=\sum_{i=1}^Nx_i^n\right\}=J^{(q,t)}_{\lambda}(\mathbf{x}).
\end{align}

The Cauchy identity for Macdonald functions is \cite{Macdonald}
\begin{eqnarray}
	\exp\left(\sum_{n=1}^{\infty}\frac{1}{n}\frac{1-t^n}{1-q^n}p_ng_n\right)=\sum_{\lambda}\frac{J^{(q,t)}_{\lambda}\{\mathbf{p}\}J^{(q,t)}_{\lambda}\{\mathbf{g}\}}{j_{\lambda}},
\end{eqnarray}
where
\begin{eqnarray}
	j_{\lambda}=\prod_{(i,j)\in\lambda} (1-q^{\lambda_i-j}t^{\lambda_j^T-i+1})(1-q^{\lambda_i-j+1}t^{\lambda_j^T-i}),
\end{eqnarray}
and $\lambda^T$ is the conjugate of $\lambda$.

The Pieri formulas for Macdonald functions are given by \cite{Macdonald,Lassalle}
\begin{subequations}\label{Pieri2}
\begin{align}
	\frac{1}{1-q}p_1J^{(q,t)}_{\lambda}\{\mathbf{p}\}&=\sum_{i\ge1} \psi_{\lambda^{(i)}/\lambda} J^{(q,t)}_{\lambda^{(i)}}\{\mathbf{p}\},\\
	\frac{1}{1-t}\frac{\partial}{\partial p_1}J^{(q,t)}_{\lambda}\{\mathbf{p}\}&=\sum_{i\ge1} \varphi_{\lambda/\lambda_{(i)}} J^{(q,t)}_{\lambda_{(i)}}\{\mathbf{p}\},
\end{align}
\end{subequations}
where $\lambda^{(i)}=(\lambda_1,\cdots,\lambda_{i-1},\lambda_i+1,\lambda_{i+1},\cdots)$ and $\lambda_{(i)}=(\lambda_1,\cdots,\lambda_{i-1},\lambda_i-1,\lambda_{i+1},\cdots)$
are partitions. The coefficients in (\ref{Pieri2}) are
\begin{subequations}\label{Pieri}
\begin{align}
	\psi_{\lambda^{(i)}/\lambda}
	&=\frac{1}{1-q}\frac{t^{l(\lambda)-i}}{1-t^{l(\lambda)}c_i}\prod_{j=1,j\neq i}^{l(\lambda)}\frac{1-t^{-1}c_j/c_i}{1-c_j/c_i},\\
	\varphi_{\lambda/\lambda_{(i)}}&=\frac{j_{\lambda}}{j_{\lambda_{(i)}}}\psi_{\lambda/\lambda_{(i)}}=t^{i-l(\lambda)}
\frac{1-t^{l(\lambda)-1}c_i}{1-q}\prod_{j=1,j\neq i}^{l(\lambda)}\frac{1-tc_j/c_i}{1-c_j/c_i},
\end{align}
\end{subequations}
where $c_i=q^{\lambda_i}t^{1-i}$ for $i=1,\cdots,N$ and $l(\lambda)=\max\{i;\lambda_i>0\}$ is the length of $\lambda$. It is easy to check that the $l(\lambda)$ in (\ref{Pieri}) can be replaced with $N$ if $N\ge l(\lambda)$.

\subsection{$(q,t)$-deformed hypergeometric functions}

We set $s,r\in\mathbb{N}$, $\mathbf{a}=(a_1,\dots,a_s)\in\mathbb{C}^s$, $\mathbf{b}=(b_1,\dots,b_r)\in\mathbb{C}^r$ with $b_j\notin\{q^{1-m}t^{n-1}|(m,n)\in \mathbb{Z}_+^2\}$
for $j=1,\cdots,r$. The Macdonald's $(q,t)$-deformed hypergeometric functions with one and two set variables are defined by \cite{MacII}
\begin{subequations}\label{hyper}
\begin{align}
	_s\Phi_r^{(q,t)}(\mathbf{a;b;x;y})=&\sum_{\lambda}\frac{\prod_{i=1}^{s}[a_i]^{(q,t)}_{\lambda}}
	{\prod_{j=1}^{r}[b_j]^{(q,t)}_{\lambda}}\frac{t^{n(\lambda)}}{J^{(q,t)}_{\lambda}(t^{\delta_N})}\frac{J^{(q,t)}_{\lambda}(\mathbf{x}) J^{(q,t)}_{\lambda}(\mathbf{y})} {j_{\lambda}},\\
	_s\phi_r^{(q,t)}(\mathbf{a;b;x})=&{_s\Phi}_r^{(q,t)}(\mathbf{a;b;x};t^{\delta_N}),
\end{align}
\end{subequations}
where $\mathbf{x}=(x_1,x_2,\cdots,x_N)$, $\mathbf{y}=(y_1,y_2,\cdots,y_N)$, $n(\lambda)=\sum_{i=1}^{l(\lambda)}(i-1) \lambda_{i}$, $t^{\delta_N}=(t^{N-1},\dots,t,1)$ and
\begin{eqnarray}\label{cij}
	[a]_{\lambda}^{(q,t)}=\prod_{(i,j)\in\lambda}(1-aq^{j-1}t^{1-i}).
\end{eqnarray}
We list some properties associated with the hypergeometric functions (\ref{hyper}) and $[a]_{\lambda}^{(q,t)}$.

(i) It is clear that $[0]_{\lambda}^{(q,t)}=1$. Then we have
\begin{eqnarray}
	_s\Phi_r^{(q,t)}(\mathbf{a;b;x;y})&=&\lim_{a\rightarrow0}{_{s+1}\Phi}_r^{(q,t)}(a,\mathbf{a;b;x;y})\nonumber\\
	&=&\lim_{b\rightarrow0}{_{s}\Phi}_{r+1}^{(q,t)}(\mathbf{a};b,\mathbf{b;x;y}).
\end{eqnarray}
In addition, the hypergeometric functions $_s\Phi_r^{(q,t)}(\mathbf{a;b;x;y})$ are invariant under the variable exchange $\mathbf{x}\leftrightarrow\mathbf{y}$;

(ii) There are some identities \cite{MacII}
\begin{subequations}\label{example}
\begin{align}
	_1\Phi_{0}^{(q,t)}(t^N;\mathbf{x;y})&=\prod_{i,j=1}^{N}\frac{(tx_iy_j;q)_{\infty}}{(x_iy_j;q)_{\infty}}=
\exp\left(\sum_{n=1}^{\infty}\frac{1-t^n}{1-q^n}\frac{p_n(\mathbf{x})p_n(\mathbf{y})}{n}\right)\nonumber\\
	&=\sum_{\lambda}\frac{J^{(q,t)}_{\lambda}(\mathbf{x}) J^{(q,t)}_{\lambda}(\mathbf{y})} {j_{\lambda}},\label{example1}\\
	{_1\phi_0^{(q,t)}}(a;\mathbf{x})&=\prod_{i=1}^{N}\frac{(ax_i;q)_{\infty}}{(x_i;q)_{\infty}}=\exp\left(\sum_{n=1}^{\infty}\frac{1-a^n}{1-q^n}\frac{p_n(\mathbf{x})}{n}\right)\nonumber\\
	&=\sum_{\lambda}\frac{J^{(q,t)}_{\lambda}(\mathbf{x})} {j_{\lambda}} J^{(q,t)}_{\lambda}\left\{p_n=\frac{1-a^n}{1-t^n}\right\},\label{example2}
\end{align}
\end{subequations}
where $(x;q)_{\infty}=\prod_{k=0}^{\infty}(1-xq^k)$ is $q$-Pochhammer symbol and $(x;q)_n=\prod_{k=0}^{n-1}(1-xq^k)$;

(iii) From (\ref{example2}) and its special case at $a=0$, we have
\begin{eqnarray}\label{aqt}
	[a]_{\lambda}^{(q,t)}=\frac{J^{(q,t)}_\lambda\left\{p_n=\frac{1-a^n}{1-t^n}\right\}}{J^{(q,t)}_\lambda\left\{p_n=\frac{1}{1-t^n}\right\}},	
\end{eqnarray}
and
\begin{eqnarray}
	J^{(q,t)}_\lambda\left\{p_n=\frac{1}{1-t^n}\right\}=\frac{J^{(q,t)}_\lambda\left\{p_n=\frac{1-t^{nN}}{1-t^n}\right\}}{[N\beta]^{(q,t)}_{\lambda}}=t^{n(\lambda)}.
\end{eqnarray}
Furthermore, by $\lim_{a\rightarrow\infty}[a]_{\lambda}^{(q,t)}/a^{|\lambda|}$ in (\ref{aqt}), we have
\begin{eqnarray}\label{csterm}
\frac{J^{(q,t)}_\lambda\left\{p_n=\frac{-1}{1-t^n}\right\}}{J^{(q,t)}_\lambda\left\{p_n=\frac{1}{1-t^n}\right\}}=(-1)^{|\lambda|}
\prod_{(i,j)\in\lambda}q^{j-1}t^{1-i}=(-1)^{|\lambda|}q^{n(\lambda^T)}t^{-n(\lambda)};
\end{eqnarray}

(iv) By (\ref{example2}) and (\ref{csterm}), we have
\begin{subequations}
\begin{align}
	\left({_0\phi_0^{(q,t)}}(\mathbf{x})\right)^{-1}&=\sum_{\lambda}\frac{J^{(q,t)}_{\lambda}(\mathbf{x})} {j_{\lambda}} J^{(q,t)}_{\lambda}\left\{p_n=\frac{-1}{1-t^n}\right\}\nonumber\\
	&=\sum_{\lambda}(-1)^{|\lambda|}q^{n(\lambda^T)}\frac{J^{(q,t)}_{\lambda}(\mathbf{x})} {j_{\lambda}},\\
	\prod_{j=1}^{s}{_1\phi}_0^{(q,t)}(a_j;\mathbf{x})&=\exp\left(\sum_{j=1}^s\sum_{n=1}^{\infty}\frac{1-a_j^n}{1-q^n}\frac{p_n(\mathbf{x})}{n}\right)\nonumber\\
	&=\sum_{\lambda}\frac{J^{(q,t)}_{\lambda}(\mathbf{x})} {j_{\lambda}} J^{(q,t)}_{\lambda}\left\{p_n=\sum_{j=1}^s\frac{1-a_j^n}{1-t^n}\right\}.
\end{align}
\end{subequations}
Note that $\left({_0\phi_0^{(q,t)}}(\mathbf{x})\right)^{-1}$ is nothing but the Kaneko's $(q,t)$-deformed hypergeometric function \cite{Kaneko2}.

Let us introduce the $(q,t)$-deformed operator $\hat{O}_{q,t}(a;\mathbf{p})$ which satisfies
\begin{eqnarray}\label{Oop}
\hat{O}_{q,t}(a;\mathbf{p})J^{(q,t)}_{\lambda}\{\mathbf{p}\}=[a]_{\lambda}^{(q,t)}J^{(q,t)}_{\lambda}\{\mathbf{p}\}.
\end{eqnarray}
Here we assume $a\notin\{q^{1-m}t^{n-1}|(m,n)\in \mathbb{Z}_+^2\}$ such that $[a]_{\lambda}^{(q,t)}\neq0$ for all $\lambda$
and $\hat{O}_{q,t}(a;\mathbf{p})$ is invertible. The operator $\hat{O}_{q,t}(a;\mathbf{p})$ has been constructed in Refs. \cite{Bourgine,Fan24,Mircomm,Fanqt}
by certain commuting subalgebra of the DIM algebra \cite{DI,Miki} or elliptic Hall algebra \cite{Schiffmann}.

In terms of $\hat{O}_{q,t}(a;\mathbf{p})$ in (\ref{Oop}), we define the $W$-operators
\begin{align}\label{wowo}
	&W_{\pm k}^{(s)}(\mathbf{a;p})= \mathbf{Ad}^{\pm1}_{\hat{O}^{(s)}_{q,t}(\mathbf{a;p})}W_{\pm k}^{(0)}(\mathbf{p}),\quad k\in\mathbb{Z}_+,\nonumber\\
	&W_k^{(0)}(\mathbf{p})=\frac{p_k}{1-q^k}, \qquad W_{-k}^{(0)}(\mathbf{p})=\frac{k}{1-t^k}\frac{\partial}{\partial p_k},
\end{align}
where we denote $\mathbf{Ad}_{\hat{e}} \hat{h}=\hat{e}\hat{h}\hat{e}^{-1}$ and $\hat{O}^{(s)}_{q,t}(\mathbf{a;p})=\prod_{i=1}^{s} \hat{O}_{q,t}(a_i;\mathbf{p})$. These $W$-operators constitute the commutative subalgebra of the DIM algebra.

Let us introduce the difference operator
\begin{eqnarray}\label{W0a}
	W_0(a;\mathbf{p})=\sum_{n=1}^{\infty}np_n\frac{\partial}{\partial p_n}-at^{1-N}\mathcal{E}_1\{\mathbf{p}\},
\end{eqnarray}
such that
\begin{eqnarray}\label{W0Mac}
	W_0(a;\mathbf{p})J^{(q,t)}_{\lambda}\{\mathbf{p}\}=\sum_{(i,j)\in\lambda}(1-aq^{j-1}t^{1-i})J^{(q,t)}_{\lambda}\{\mathbf{p}\}.
\end{eqnarray}
It should be mentioned that the spectrum of the operator $W_0(a;\mathbf{x})$ is more compatible with the factor $[a]_{\lambda}^{(q,t)}$ in (\ref{hyper})
than the $W$-operators defined in \cite{Fanqt}.

By the Peiri formulas (\ref{Pieri}), it is easy to check that the $W$-operators (\ref{wowo}) with $k=1$ can be given by the nested commutators \cite{Fan24}
\begin{eqnarray}\label{W0op}
	W_{\pm1}^{(s)}(\mathbf{a;p})=\prod_{j=1}^{s}\mathbf{ad}_{\pm W_0(a_j;\mathbf{p})}W_{\pm1}^{(0)}(\mathbf{p}),
\end{eqnarray}
where $\mathbf{ad}_{\hat{e}} \hat{h}=[\hat{e},\hat{h}]$.
Then we have
\begin{subequations}
\begin{align}\label{EW0}
	W_{-1}^{(1)} (t^N;\mathbf{p})&=\left[\frac{1}{1-t}\frac{\partial}{\partial p_1},W_0(t^N;\mathbf{p})\right]=\mathcal{E}_0\{\mathbf{p}\},\\
	W_1^{(1)}(a;\mathbf{p})&=\left[W_0(a;\mathbf{p}),\frac{p_1}{1-q}\right]=t^{1-N}a\mathcal{E}_2\{\mathbf{p}\}+\frac{1-a}{1-q}p_1.
\end{align}
\end{subequations}
Based on (\ref{ekpnx}), (\ref{W0a}) and (\ref{W0op}), some $W$-operators (\ref{wowo}) can be expressed in terms of the variables $\mathbf{x}$, i.e.
\begin{align}
	W_0(a;\mathbf{x})&=W_0(a;\mathbf{p})\Big|_{p_n=\sum_{i=1}^Nx_i^n},\nonumber\\
	W_{1}^{(s)}(\mathbf{a;x})&= W_{ 1}^{(s)}(a;\mathbf{p})\Big|_{p_n=\sum_{i=1}^Nx_i^n},\nonumber\\
	W_{-1}^{(r+1)}(t^N,\mathbf{b;x})&= W_{ -1}^{(r+1)}(t^N,\mathbf{b;p})\Big|_{p_n=\sum_{i=1}^Nx_i^n}.
\end{align}
Note that $\mathcal{E}_0(\mathbf{x})$ can also be rewritten as \cite{Lassalle}
\begin{eqnarray}
	\mathcal{E}_0(\mathbf{x})=\left[\frac{1}{1-q}\sum_{i=1}^{N}x_i^{-1},W_0(qt^{N-1};\mathbf{x})\right].
\end{eqnarray}

For convenience, we denote
\begin{align}
	\hat{O}^{(s)}_{q,t}(\mathbf{a;x})=\hat{O}^{(s)}_{q,t}(\mathbf{a;p})\Big|_{p_n=\sum_{i=1}^Nx_i^n},\nonumber\\
	\hat{O}^{(r)}_{q,t}(\mathbf{b;y})=\hat{O}^{(r)}_{q,t}(\mathbf{b;p})\Big|_{p_n=\sum_{i=1}^Ny_i^n}.
\end{align}
Then the hypergeometric functions (\ref{hyper}) can be represented as
\begin{subequations}\label{orep}
\begin{align}
	{_s\Phi}_r^{(q,t)}(\mathbf{a;b;x;y})&=\hat{O}^{(s)}_{q,t}(\mathbf{a;x})\left(\hat{O}^{(r)}_{q,t}(\mathbf{b;y})\right)^{-1}{_0\Phi}_0^{(q,t)}(\mathbf{x;y}),\\
	_s\phi_r^{(q,t)}(\mathbf{a;b;x})&=\hat{O}^{(s)}_{q,t}(\mathbf{a;x})\left(\hat{O}^{(r)}_{q,t}(\mathbf{b;x})\right)^{-1}{_0\phi}_0^{(q,t)}(\mathbf{x}).
\end{align}
\end{subequations}
Due to (\ref{example}) and (\ref{wowo}), we obtain
\begin{subequations}\label{wrep}
\begin{align}
	&{_{s+1}\Phi}_0^{(q,t)}(t^N,\mathbf{a;x;y})
	\nonumber\\
	=&\exp\left(\sum_{n=1}^{\infty}(1-t^n)(1-q^n)\frac{W_n^{(s_1)}(\mathbf{\hat{a};x})W_n^{(s-s_1)}(\mathbf{\check{a};y})}{n}\right)\cdot1, \\
	&{_{s+1}\phi}_0^{(q,t)}(a,\mathbf{a;x})
	\nonumber\\
	=&\exp\left(\sum_{n=1}^{\infty}(1-a^n)\frac{W_n^{(s)}(\mathbf{a;x})}{n}\right)\cdot1\nonumber\\
	=&\exp\left(\sum_{n=1}^{\infty}\frac{W_n^{(s+1)}(a,\mathbf{a;x})}{n}\right)\cdot1,
\end{align}
\end{subequations}
where $0 \le s_1\le s$, $\mathbf{\hat{a}}=(a_{i_1},a_{i_2},\cdots,a_{i_{s_1}})$, $\mathbf{\check{a}}=(a_{j_1},a_{i_2}, \cdots,a_{j_{s-s_1}})$
and $\{i_k;1\le k\le s_1\}\sqcup\{j_k;1\le k\le s-s_1\}=\{1,2,\cdots,s\}$.
We call (\ref{wrep}) the $W$-representations of the hypergeometric functions (\ref{hyper}) with $r=0$.
It is clear that these $W$-representations are not unique.

\subsection{Constraints of $(q,t)$-deformed hypergeometric functions}
From (\ref{example1}), it is clear that
\begin{eqnarray}\label{W+W-}
\left(\frac{p_1(\mathbf
x)}{1-q}-\frac{1}{1-t}\frac{\partial}{\partial p_1(\mathbf{y})}\right){_1\Phi_0^{(q,t)}(t^N;\mathbf{x;y})}=0.
\end{eqnarray}
By taking the action of
$\hat{O}^{(s+l)}_{q,t}(\mathbf{a,d;\mathbf{x}})\left(\hat{O}_{q,t}(t^N;\mathbf{y})\hat{O}^{(r+l)}_{q,t}(\mathbf{b,d;y})\right)^{-1}$
on (\ref{W+W-}), we obtain that the hypergeometric functions ${_s\Phi_r^{(q,t)}(\mathbf{a;b;x;y})}$ in (\ref{hyper}) satisfy
\begin{eqnarray}\label{ConsXY}
\left(W_1^{(s+l)}(\mathbf{a,d;x})-W_{-1}^{(r+l+1)}(t^N,\mathbf{b,d;y})\right){_s\Phi_r^{(q,t)}(\mathbf{a;b;x;y})}=0,
\end{eqnarray}
where $l\in\mathbb{N}$ and $\mathbf{d}=(d_1,\dots,d_l)$.
In addition, (\ref{ConsXY}) still holds by taking the variable exchange $\mathbf{x}\leftrightarrow\mathbf{y}$.

Taking $(s,r,l)=(0,0,1)$ in (\ref{ConsXY}), it gives
\begin{eqnarray}\label{E200}
	\left(W_1^{(1)}(d_1;\mathbf{x})-W_{-1}^{(2)}(t^N,d_1;\mathbf{y})\right){_0\Phi_0^{(q,t)}(\mathbf{x,y})}=0,
\end{eqnarray}
where
\begin{eqnarray}
	W_1^{(1)}(d;\mathbf{x})=\frac{1-d}{1-q}p_1(\mathbf{x})+dt^{1-N}\mathcal{E}_2(\mathbf{x}).
\end{eqnarray}
In addition, it is easy to check that
\begin{eqnarray}\label{E1rs}
	\mathcal{E}_1(\mathbf{x}){_s\Phi_r^{(q,t)}}(\mathbf{a;b;x;y})=\mathcal{E}_1(\mathbf{y}){_s\Phi_r^{(q,t)}(\mathbf{a;b;x;y})}.
\end{eqnarray}
It should be mentioned that (\ref{E200}) and (\ref{E1rs}) will be used to derive the constraints of certain $(q,t)$-deformed matrix models in next section.

Before presenting the constraints of ${_s\phi_r^{(q,t)}}(\mathbf{a;b;x})$ in (\ref{hyper}), let us define some combination numbers related to $\lambda$
\begin{subequations}
\begin{align}
	K_{l,n}(\lambda)&:=\sum_{i=1}^n(c_i)^l\psi_{\lambda^{(i)}/\lambda}t^{i-1}(1-t^nc_i)\nonumber\\
	&=\frac{t^{n-1}}{1-q}\sum_{i=1}^n(c_i)^l\prod_{j=1,j\neq i}^{n}\frac{1-t^{-1}c_j/c_i}{1-c_j/c_i},\\
	\bar{K}_{l,n}(\lambda)&:=\sum_{i=1}^n\varphi_{\lambda/\lambda_{(i)}}t^{1-i} \frac{(q^{-1}c_i)^l}{1-t^{n-1}c_i}\nonumber\\
	&=\frac{t^{1-n}}{1-q}\sum_{i=1}^n(q^{-1}c_i)^l\prod_{j=1,j\neq i}^{n}\frac{1-tc_j/c_i}{1-c_j/c_i},
\end{align}
\end{subequations}
for $l\in\mathbb{N}$ and $n\ge l(\lambda)$, where $c_i=q^{\lambda}t^{1-i}$, $\psi_{\lambda^{(i)}/\lambda}$ and $\varphi_{\lambda/\lambda_{(i)}}$ come from (\ref{Pieri}).
Then their generating functions are given by
\begin{subequations}\label{KZKZbar}
\begin{align}
	K_n(\lambda;z):=&\sum_{l=0}^{\infty}K_{l,n}(\lambda)z^l=\frac{t^{n-1}}{1-q}\sum_{i=1}^n\frac{1}{1-c_iz}\prod_{j=1,j\neq i}^{n}\frac{1-t^{-1}c_j/c_i}{1-c_j/c_i}\nonumber\\
	=&\frac{t^{n-1}}{(1-q)(1-t^{-1})}\left(\prod_{j=1}^{n}\frac{1-t^{-1}c_jz}{1-c_jz}-t^{-n}\right),\\
	\bar{K}_n(\lambda;z):=&\sum_{l=0}^{\infty}\bar{K}_{l,n}(\lambda)z^l=\frac{t^{1-n}}{1-q}\sum_{i=1}^n\frac{1}{1-q^{-1}c_iz}\prod_{j=1,j\neq i}^{n}\frac{1-tc_j/c_i} {1-c_j/c_i}\nonumber\\
	=&\frac{t^{1-n}}{(1-q)(1-t)}\left(\prod_{j=1}^{n}\frac{1-q^{-1}tc_jz}{1-q^{-1}c_jz}-t^n\right).
\end{align}
\end{subequations}
The equations (\ref{KZKZbar}) can be checked by taking the residue theorem at the singular points $z=\infty$ and $z=c_i^{-1}$,  $i=1,2,\cdots,n$,
for $K_{n}(\lambda;z)$ or $z=qc_i^{-1}$ for $\bar{K}_{n}(\lambda;z)$.

Moreover, we denote the functions
\begin{subequations}\label{KaKb}
\begin{align}
	K_n^{(s)}(\lambda;\mathbf{a})&:=\sum_{i=1}^n\psi_{\lambda^{(i)}/\lambda}t^{i-1}(1-t^nc_i)\prod_{j=1}^{s}(1-a_jc_i)\nonumber\\
	&=\prod_{j=1}^sa_j\sum_{l=0}^s(-1)^le_{s-l}(a_1^{-1},\cdots,a_s^{-1})K_{l,n}(\lambda),\label{qKra}\\
	\bar{K}_n^{(r)}(\lambda;\mathbf{b})&:=\sum_{i=1}^nt^{1-i}\varphi_{\lambda/\lambda_{(i)}} \frac{\prod_{k=1}^{r} (1-b_kq^{-1}c_i)}{1-t^{n-1}c_i}\nonumber\\
	&=\prod_{k=1}^rb_k\sum_{l=0}^r(-1)^le_{r-l}(b_1^{-1},\cdots,b_r^{-1})\bar{K}_{l,n}(\lambda),
\end{align}
\end{subequations}
where $a_j,b_k\neq 0$, $n\in\mathbb{Z}_+$, $e_l$ are elementary symmetric polynomials which satisfy $\prod_{j=1}^{s}(1+a_jx)=\sum_{l=1}^{s}e_l(a_1,\cdots,a_s)x^{s-l}$.

A straightforward calculation shows that
\begin{eqnarray}\label{W-K}
	&&W^{(1)}_{-1}(t^N;\mathbf{x}){_s\phi^{(q,t)}_0}(\mathbf{a;x})\nonumber\\
	&=&\left(W^{(s)}_{1}(\mathbf{a;y}){_s\Phi^{(q,t)}_0}(\mathbf{a;x;y})\right)\Big|_{\mathbf{y}=t^{\delta_N}}\nonumber\\
	&=&\hat{O}_{q,t}^{(s)}(\mathbf{a;y})\left(\frac{p_1(\mathbf{y})}{1-q}{_0\Phi^{(q,t)}_0(\mathbf{x;y})}\right)\Big|_{\mathbf{y}=t^{\delta_N}}\nonumber\\
	&=&\sum_{\lambda}\sum_{i=1}^N \psi_{\lambda^{(i)}/\lambda}\prod_{j=1}^{s}[a_j]^{(q,t)}_{\lambda^{(i)}} \frac{J^{(q,t)}_{\lambda^{(i)}}(t^{\delta_N})}
    {J^{(q,t)}_{\lambda}(t^{\delta_N})}\frac{t^{n(\lambda)}J^{(q,t)}_{\lambda} (\mathbf{x})}{j_\lambda}\nonumber\\
	&=&\sum_{\lambda}K_N^{(s)}(\lambda;\mathbf{a}) \prod_{j=1}^{s}[a_j]^{(q,t)}_{\lambda} \frac{t^{n(\lambda)}J^{(q,t)}_{\lambda}(\mathbf{x})}{j_{\lambda}},
\end{eqnarray}
and
\begin{eqnarray}\label{W+K}
	&&W^{(0)}_{1}(\mathbf{x}){_1\phi^{(q,t)}_r}(qt^{N-1};\mathbf{b;x})\nonumber\\&=&\sum_{\lambda}\sum_{i=1}^N\psi_{\lambda^{(i)}/\lambda}
\frac{[qt^{N-1}]^{(q,t)}_{\lambda}} {\prod_{k=1}^{r}[b_k]^{(q,t)}_{\lambda}} \frac{t^{n(\lambda)} J^{(q,t)}_{\lambda^{(i)}} (\mathbf{x})}{j_\lambda} \nonumber\\
	&=&\sum_{\lambda}\sum_{i=1}^N \psi_{\lambda/\lambda_{(i)}} \frac{[qt^{N-1}]^{(q,t)}_{\lambda_{(i)}}}{\prod_{k=1}^{r}[b_k]^{(q,t)}_{\lambda_{(i)}}}
\frac{t^{n(\lambda_{(i)})} J^{(q,t)}_{\lambda} (\mathbf{x})}{j_{\lambda_{(i)}}} \nonumber\\
	&=&\sum_{\lambda}\bar{K}_N^{(r)}(\lambda;\mathbf{b})\frac{[qt^{N-1}]^{(q,t)}_{\lambda}}{\prod_{k=1}^{r}[b_k]^{(q,t)}_{\lambda}}
\frac{t^{n(\lambda)}J^{(q,t)}_{\lambda}(\mathbf{x})}{j_{\lambda}}.
\end{eqnarray}

In terms of Macdonald's difference operators (\ref{Dqt}), we define
\begin{subequations}\label{WsWr}
\begin{align}
	\mathcal{W}^{(s)}_0(\mathbf{a;x})=&\frac{t^{N-1}}{(1-q)(1-t^{-1})}\prod_{j=1}^sa_j\sum_{l=0}^s(-1)^le_{s-l}(a_1^{-1},\cdots,a_s^{-1})\nonumber\\
	&\times\oint\frac{dz}{2\pi\mathbf{i}z^{1+j}}\left(\frac{\mathcal{D}_N(-t^{-N}z;\mathbf{x})}{\mathcal{D}_N(-t^{1-N}z;\mathbf{x})}-t^{-N}\right),\\
	\mathcal{\bar{W}}^{(r)}_0(\mathbf{b;x})=&\frac{t^{1-N}}{(1-q)(1-t)}\prod_{k=1}^rb_k\sum_{l=0}^r(-1)^le_{r-l}(b_1^{-1},\cdots,b_r^{-1})\nonumber\\
	&\times\oint\frac{dz}{2\pi\mathbf{i}z^{1+k}}\left(\frac{\mathcal{D}_N(-q^{-1}t^{2-N}z;\mathbf{x})}{\mathcal{D}_N(-q^{-1}t^{1-N}z;\mathbf{x})}-t^{N}\right),
\end{align}
\end{subequations}
such that
\begin{subequations}\label{W0eigen}
\begin{align}
	\mathcal{W}^{(s)}_0(\mathbf{a;x})J_{\lambda}^{(q,t)}(\mathbf{x})&=K_N^{(s)}(\lambda;\mathbf{a})J_{\lambda}^{(q,t)}(\mathbf{x}),\\
	\mathcal{\bar{W}}^{(r)}_0(\mathbf{b;x})J_{\lambda}^{(q,t)}(\mathbf{x})&=\bar{K}_N^{(r)}(\lambda;\mathbf{b})J_{\lambda}^{(q,t)}(\mathbf{x}).
\end{align}
\end{subequations}
Here the parameter $N$, as the number of variables $x_i$, will change when there is a variable substitution.

It follows from (\ref{W-K}), (\ref{W+K}) and (\ref{W0eigen}) that
\begin{subequations}
\begin{align}
    \left(W^{(1)}_{-1}(t^N,\mathbf{x})-\mathcal{W}_0^{(s)}(\mathbf{a;x})\right){_s\phi_0^{(q,t)}}(\mathbf{a;x})&=0,\label{W-W0}\\
    \left(W^{(0)}_{1}(\mathbf{x})-\mathcal{\bar{W}}^{(r+1)}_0(qt^{N-1},\mathbf{b;x})\right){_0\phi^{(q,t)}_r}(\mathbf{b;x})&=0,\label{W+W0}
\end{align}
\end{subequations}
where we have replaced $\mathbf{b}=(b_1,\cdots,b_r)$ with $(qt^{N-1},b_1,\cdots,b_r)$ in (\ref{W+W0}).

Taking the actions of $(\hat{O}_{q,t}^{(r)}(\mathbf{b;x}))^{-1}$ on (\ref{W-W0}) and $\hat{O}_{q,t}^{(s)}(\mathbf{a;x})$ on (\ref{W+W0}), we obtain the constraints
\begin{subequations}\label{ConsW+-0}
\begin{align}
	\left(W^{(r+1)}_{-1}(t^N,\mathbf{b;x})-\mathcal{W}_0^{(s)}(\mathbf{a;x})\right){_s\phi_r^{(q,t)}}(\mathbf{a;b;x})&=0,\\
	\left(W^{(s)}_{1}(\mathbf{a;x})-\mathcal{\bar{W}}^{(r+1)}_0(qt^{N-1},\mathbf{b;x})\right){_s\phi^{(q,t)}_r}(\mathbf{a;b;x})&=0.
\end{align}
\end{subequations}

Taking $\mathbf{a}=a$ and $\mathbf{b}=qt^{N-1}$ in (\ref{WsWr}), it gives
\begin{align}\label{mcalW}
	\mathcal{W}^{(1)}_0(a;\mathbf{x})&=a\mathcal{E}_1(\mathbf{x})+\frac{1-a}{1-q}\{N\}_t,\nonumber\\
	\mathcal{\bar{W}}^{(1)}_0(qt^{N-1};\mathbf{x})&=t^{1-N}\mathcal{E}_1(\mathbf{x}).
\end{align}
Then by (\ref{ConsW+-0}), we obtain
\begin{subequations}\label{LsLr}
\begin{align}
	\mathcal{L}_{(1,r)}(a;\mathbf{b};w;\mathbf{x}){_1\phi}_r^{(q,t)}(a;\mathbf{b};w\mathbf{x})=&0,\label{L1rcons}\\
	\bar{\mathcal{L}}_{(s,0)}(\mathbf{a};w;\mathbf{x}){_s\phi}_0^{(q,t)}(\mathbf{a};w\mathbf{x})=&0,
\end{align}
\end{subequations}
where we denote
\begin{subequations}
\begin{align}
	\mathcal{L}_{(1,r)}(a;\mathbf{b};w;\mathbf{x})=&w^{-1}W_{-1}^{(r+1)}(t^N,\mathbf{b;x})-\mathcal{W}^{(1)}_0(a;\mathbf{x}),\label{L1r}\\
	\bar{\mathcal{L}}_{(s,0)}(\mathbf{a};w;\mathbf{x})
	=&wW_1^{(s)}(\mathbf{a;x})-\mathcal{\bar{W}}^{(1)}_0(qt^{N-1};\mathbf{x}).
\end{align}
\end{subequations}

Let us turn to consider the product
\begin{align}
	\Psi(\mathbf{a;w;x})&={_1\phi_0^{(q,t)}}(a_1;w_1\mathbf{x}){_1\phi_0^{(q,t)}}(a_2;w_2\mathbf{x})\nonumber\\
	&=\exp\left(\sum_{n=1}^{\infty}\frac{w_1^n(1-a_1^n)+w_2^n(1-a_2^n)}{1-q^n}\frac{p_n(\mathbf{x})}{n}\right),
\end{align}
where $\mathbf{a}=(a_1,a_2)$ and $\mathbf{w}=(w_1,w_2)$.

By (\ref{Ecoll}), we have
\begin{align}\label{calcu}
	&\Psi^{-1}(\mathbf{a;w;x})\left(\mathcal{E}_k(\mathbf{x})\Psi(\mathbf{a;w;x})\right)\nonumber\\
	=&\frac{1}{(1-q)(t-1)}\oint\frac{dz}{2\pi \mathbf{i}} z^{-k}\left[t^N\exp\left(\sum_{n=1}^{\infty} \frac{1-t^{-n}}{n}p_n(\mathbf{x})z^n\right)-1\right]\nonumber\\
	&\times\left[1-\frac{(1-w_1z^{-1})(1-w_2z^{-1})}{(1-a_1w_1z^{-1})(1-a_2w_2z^{-1})}\right].
\end{align}
We denote
\begin{align}\label{L11k}
	\mathcal{L}_k(\mathbf{a,w;x})=&\mathcal{E}_k(\mathbf{x})-(a_1w_1+a_2w_2)\mathcal{E}_{k+1}(\mathbf{x})\nonumber\\
	&+(a_1a_2w_1w_2)\mathcal{E}_{k+2}(\mathbf{x})+(1-a_1a_2)w_1w_2\mathcal{A}_{k+1}(\mathbf{x})\nonumber\\
	&-[(1-a_1)w_1+(1-a_2)w_2]\mathcal{A}_k(\mathbf{x}),\qquad k\ge0,
\end{align}
such that
\begin{eqnarray}\label{L1010}
	\mathcal{L}_k(\mathbf{a,w;x})\Psi(\mathbf{a;w;x})=0.
\end{eqnarray}
Taking $k=0$ and $1$ in (\ref{L11k}), it gives
\begin{subequations}
\begin{align}
	\mathcal{L}_0(\mathbf{a,w;x})=&W^{(1)}_{-1}(t^N;\mathbf{x})-[(1-a_1)w_1+(1-a_2)w_2]\frac{\{N\}_t}{1-q}\nonumber\\
	&-(a_1w_1+a_2w_2)\mathcal{E}_1(\mathbf{x})+w_1w_2t^{N-1}W_1^{(1)}(a_1a_2;\mathbf{x}),\\
	\mathcal{L}_1(\mathbf{a,w;x})=&\mathcal{E}_1(\mathbf{x})-(w_1+w_2)t^{N-1}W_1^{(1)}\left(\frac{a_1w_1+a_2w_2}{w_1+w_2};\mathbf{x}\right)\nonumber\\
	&+w_1w_2[a_1a_2\mathcal{E}_3(\mathbf{x})+(1-a_1a_2)\mathcal{A}_2(\mathbf{x})].
\end{align}
\end{subequations}

Let us denote
\begin{subequations}\label{L10k}
\begin{align}
	L_k(a_1,w_1;\mathbf{x})&:=\lim_{w_2\rightarrow0}\mathcal{L}_k(\mathbf{a,w;x}),\\
	\tilde{L}_k(w;\mathbf{x})&:=\lim_{a\rightarrow\infty}L_k(a,w/a;\mathbf{x}),
\end{align}
\end{subequations}
such that
\begin{subequations}\label{degrad}
\begin{align}
	L_k(a_1,w_1;\mathbf{x}){_1\phi}_0^{(q,t)}(a_1;w_1\mathbf{x})=0,\\
	\tilde{L}_k(w;\mathbf{x})\left({_0\phi}_0^{(q,t)}(w\mathbf{x})\right)^{-1}=0.
\end{align}
\end{subequations}
We list some operators in (\ref{L10k})
\begin{subequations}
\begin{align}
	L_0(a,w;\mathbf{x})&=w\mathcal{L}_{(1,0)}(a;w;\mathbf{x})=
	W_{-1}^{(1)}(t^N;\mathbf{x})- w\left(a\mathcal{E}_1(\mathbf{x})+\frac{1-a}{1-q}\{N\}_t\right),\label{L01cons}\\
	L_1(a,w;\mathbf{x})
	&=-t^{N-1}\bar{\mathcal{L}}_{(1,0)}(a;w;\mathbf{x})=\mathcal{E}_1(\mathbf{x})-wt^{N-1}W_1^{(1)}(a;\mathbf{x}),\\
	\tilde{L}_0(w;\mathbf{x})
	&=W_{-1}^{(1)}(t^N;\mathbf{x})- w\mathcal{E}_1(\mathbf{x})+w\frac{1}{1-q}\{N\}_t.\label{L00cons}
\end{align}
\end{subequations}

\begin{theorem}\label{UniqueL11}
$\Psi(\mathbf{a;w;x})$ is the unique symmetric function solution of the constraints
\begin{eqnarray}
	\mathcal{L}_0(\mathbf{a,w;x})F(\mathbf{x})=0,
\end{eqnarray}
subject to the initial condition $F(0^N)=1$ and the stability condition that for every $1\le n\le N$, $\Psi(\mathbf{a;w;x}_n)$ is a solution of the constraint
\begin{eqnarray}
	\mathcal{L}_0(\mathbf{a,w};\mathbf{x}_n)F(\mathbf{x}_n)=0,
\end{eqnarray}
where $\mathbf{x}_n=(x_1,x_2,\cdots,x_n)$ and $\mathbf{x}=\mathbf{x}_N$.
\end{theorem}
\begin{proof}
Let us denote $S(\mathbf{x})=F(\mathbf{x})\Psi^{-1}(\mathbf{a;w;x})$.
A straightforward calculation gives
\begin{align}
	\mathcal{B}_k(\mathbf{x}):=&\mathbf{Ad}^{-1}_{\Psi(\mathbf{a;w;x})}\mathcal{L}_k(\mathbf{a,w;x})\nonumber\\
	=&\mathcal{E}_k(\mathbf{x}) -(w_1+w_2)\mathcal{E}_{k+1}(\mathbf{x}) +w_1w_2\mathcal{E}_{k+2}(\mathbf{x}).
\end{align}
Then we only need to consider the difference system
\begin{eqnarray}\label{BSN}
	\mathcal{B}_0(\mathbf{x}_n)S(\mathbf{x}_n)=0,\qquad 1\le n\le N.
\end{eqnarray}
We denote	$S(\mathbf{x}_n)=\sum_{\lambda}\gamma_{\lambda}J^{(q,t)}_{\lambda}(\mathbf{x}_n)$.
It follows from (\ref{Pieri2}) and (\ref{BSN}), we obtain
\begin{align}\label{BSM1}
&\sum_{i=1}^n\gamma_{\lambda^{(i)}}\varphi_{\lambda^{(i)}/\lambda}\left(1-q^{\lambda_i}t^{n+1-i}\right)-(w_1+w_2)
\gamma_{\lambda}\sum_{(i,j)\in\lambda}q^{j-1}t^{n-i}\nonumber\\
+&w_1w_2\sum_{i=1}^n\gamma_{\lambda_{(i)}}\psi_{\lambda/\lambda_{(i)}}\left(1-q^{\lambda_i-1}t^{1-i}\right)=0,
\end{align}
for $1\le n\le N$.

Comparing the coefficient of $t^n$ and constant terms in (\ref{BSM1}), it gives
\begin{subequations}\label{BSM2}
\begin{align}
	\sum_{i=1}^n\gamma_{\lambda^{(i)}}\psi_{\lambda^{(i)}/\lambda}+w_1w_2\sum_{i=1}^n\gamma_{\lambda_{(i)}}\psi_{\lambda/\lambda_{(i)}}\left(1-q^{\lambda_i-1}t^{1-i}\right)&=0,\\
	\sum_{i=1}^n\gamma_{\lambda^{(i)}}\psi_{\lambda^{(i)}/\lambda}q^{\lambda_i}t^{1-i}-(w_1+w_2)\gamma_{\lambda}\sum_{(i,j)\in\lambda}q^{j-1}t^{-i}&=0.
\end{align}
\end{subequations}
Let $P(d)$ be the number of partitions of size $d$. The number of unknown $\gamma_{\lambda^{(i)}}$ $P(d+1)$ is no more than the number of equations $2P(d)$.
Thus (\ref{BSM2}) is an over-determined linear system of equations. Furthermore, it is easy to check that $\gamma_{\lambda^{(i)}}=0$
for all partition $\lambda$, i.e., $S(\mathbf{x})\equiv1$.
Therefore, we obtain  the unique solution $F(\mathbf{x})={_1\phi_0^{(q,t)}}(a_1;w_1\mathbf{x}){_1\phi_0^{(q,t)}}(a_2;w_2\mathbf{x})$.
\end{proof}
By similar proofs, we conclude that the hypergeometric functions ${_s\phi^{(q,t)}_r}(\mathbf{a;b;x})$ are respectively the unique solutions of the constraints
\begin{subequations}
\begin{align}
	\left(W^{(r+1)}_{-1}(t^N,\mathbf{b;x})-\mathcal{W}_0^{(s)}(\mathbf{a;x})\right)F(\mathbf{x})&=0,\\
	\left(W^{(s)}_{1}(\mathbf{a;x})-\mathcal{\bar{W}}^{(r+1)}_0(qt^{N-1},\mathbf{b;x})\right)\bar{F}(\mathbf{x})&=0,
\end{align}
\end{subequations}
in the sense of the corresponding initial conditions and stability conditions.

We call (\ref{ConsW+-0}) and (\ref{L1010}) with $k=0$ the hypergeometric constraints where the constraint operators are linearly composed of several homogeneous terms
with degree $-1,0$ and $1$. These constraints will play an important role in investigating the relations between hypergeometric functions (\ref{hyper})
and certain $(q,t)$-deformed matrix models in next section.

\section{Some $(q,t)$-analogues of matrix models}

\subsection{The refined Chern-Simons matrix model}
Let us start from the matrix model description of Chern-Simons theory with level $k$ defined on a three-sphere $\mathbf{S}^3$ \cite{Beasley}.
When we take the gauge group $G=U(N)$, the partition function is given by the Stieltjes-Wigert matrix integral \cite{Tierz,Marino}
\begin{align}\label{CS}
Z^{CS}=&\frac{1}{N!}\int_{\mathbb{R}^N_{+}}d\mathbf{x}\prod_{1\le i\neq j\le N}(1-x_i/x_j)\prod_{i=1}^{N}x_i^{-1} e^{-\frac{\log^2x_i}{2\log q}},
\end{align}
where $q=\exp\left(\frac{2\pi\mathbf{i}}{k+ N}\right)$.

We denote the Schur polynomials $s_{\lambda}(\mathbf{x})=J^{(q,q)}_{\lambda}(\mathbf{x})$.
Their normalized averages may provide a class of unknot invariants in $\mathbf{S}^3$.
The superintegrability relation is \cite{Tierz}
\begin{align}
\langle s_{\lambda}(\mathbf{x}) \rangle^{(CS)}=&\frac{1}{Z^{CS}}\int_{\mathbb{R}^N_{+}}d\mathbf{x}\prod_{1\le i\neq j\le N}(1-x_i/x_j)\prod_{i=1}^{N}x_i^{-1}
e^{-\frac{\log^2x_i}{2\log q}} s_{\lambda}(\mathbf{x})\nonumber\\
=&q^{\frac{1}{2}\sum_{i=1}^N(\lambda_i-2i)\lambda_i}s_{\lambda}\left\{p_k=\frac{1-q^{kN}}{1-q^k}\right\}.
\end{align}
More similar results associated with the supersymmetric gauge group $U(N|M)$ see \cite{Eynard}.

By using refined Chern-Simons theory and knot homology, one can construct the partition function of refined Chern-Simons model \cite{Dunfield,Aganagic,Aganagic2}
by a specific type of deformation of the matrix integral (\ref{CS}).
Then for the unknot refined Chern-Simons model, the partition function can be written as the $q$-integral
\begin{eqnarray}\label{rCS}
Z^{rCS}(a)=\int_{\mathbb{R}_+^N}d_q\mathbf{x}\prod_{i=1}^{N} x_i^ae^{-\frac{\log^2x_i}{2\log q}} \Delta_{q,t}(\mathbf{x}),
\end{eqnarray}
where $\Delta_{q,t}(\mathbf{x})=\prod_{1\le i\neq j\le N}\prod_{k=0}^{\infty} \frac{1-q^kx_i/x_j}{1-q^ktx_i/x_j}$ is the $(q,t)$-deformed Vandermonde determinant
and $d_q\mathbf{x}$ is the $q$-measure defined by (\ref{qmeasure}).
Note that the refined partition function is defined by the ordinary measure $d\mathbf{x}$ in \cite{Cassia22}.

The refined gauge theory requires that the parameters $q$,$t$ and $a$ have the following forms
\begin{eqnarray}
	q=\exp\left(\frac{2\pi\mathbf{i}}{k+\beta N}\right),\quad t=\exp\left(\frac{2\pi\mathbf{i}\beta}{k+\beta N}\right),\quad
	a=(N-1)\beta-N.
\end{eqnarray}
However, the matrix integral (\ref{rCS}) makes sense for arbitrary parameters $q$,$t$ and $a$ with $t=q^{\beta}\in(0,1)$ and $\operatorname{Re}a\ge-1$.

We denote $Z^{rCS}=Z^{rCS}(a)$ and the normalized average of the integral (\ref{rCS}) for arbitrary symmetric function $f(\mathbf{x})$ is
\begin{eqnarray}\label{avf}
	\langle f(\mathbf{x})\rangle^{(rCS)}= \frac{1}{Z^{CS}} \int_{\mathbb{R}_+^N}d_q\mathbf{x} \prod_{i=1}^{N} x_i^ae^{-\frac{\log^2x_i}{2\log q}} \Delta_{q,t}(\mathbf{x})f(\mathbf{x}).
\end{eqnarray}

The conventional scheme for calculating the average $\langle J^{(q,t)}_{\lambda}\rangle$ is to consider the generating function
\begin{align}\label{tauCS}
    &\langle_1\Phi_0(t^N;\mathbf{x;y})\rangle^{(rCS)}\nonumber\\
    =&\int_{\mathbb{R}_+^N}d_q\mathbf{x}\prod_{i=1}^{N} x_i^ae^{-\frac{\log^2x_i}{2\log q}} \Delta_{q,t}(\mathbf{x})\exp\left(\sum_{k=1}^{\infty}\frac{1-t^k}{1-q^k} \frac{p_k(\mathbf{x})p_k(\mathbf{y})}{k}\right)
\end{align}
and its $(q,t)$-deformed Virasoro constraints.

The normalized averages $\langle J_{\lambda}^{(q,t)}(\mathbf{x}) \rangle^{(rCS)}$ may provide certain refined unknot invariants and were conjectured in \cite{Cassia22}
\begin{align}\label{avCS}
	\langle J^{(q,t)}_{\lambda}(\mathbf{x}) \rangle^{(rCS)}=&[t^N]^{(q,t)}_{\lambda} J^{(q,t)}_\lambda\left\{p_k=-\frac{(-q^{a+3/2})^k}{1-t^k}\right\}\nonumber\\
	=&\frac{J^{(q,t)}_\lambda\left\{p_k=\frac{1-t^{Nk}}{1-t^k}\right\}}{J^{(q,t)}_\lambda\left\{p_k=\frac{1}{1-t^k}\right\}}J^{(q,t)}_\lambda\left\{p_k=-\frac{(-q^{a+3/2})^k}{1-t^k}\right\}.
\end{align}
Let us give a short proof for it.

\textit{Proof of (\ref{avCS})}: Let us consider the normalized average
\begin{align}\label{00CS}
	\langle _0\Phi_0^{(q,t)}(\mathbf{x,y}) \rangle^{(rCS)}&=\frac{1}{Z^{rCS}} \int_{\mathbb{R}_+^N}d_q\mathbf{x} \prod_{i=1}^{N} x_i^ae^{-\frac{\log^2x_i}{2\log q}}
    \Delta_{q,t}(\mathbf{x})_0 \Phi_0^{(q,t)}(\mathbf{x,y})\nonumber\\
    &=\sum_{\lambda}\frac{J^{(q,t)}_{\lambda}(\mathbf{y})}{[t^{N}]_{\lambda}^{(q,t)}j_{\lambda}}\langle J^{(q,t)}_{\lambda}(\mathbf{x}) \rangle^{(rCS)},
\end{align}
which is the simplest generating function of $\langle J^{(q,t)}_{\lambda}(\mathbf{x}) \rangle^{(rCS)}$.

We introduce the total derivative operator
\begin{eqnarray}
	\sum_{i=1}^{N}\frac{\partial}{\partial_qx_i}A_{t^{-1},i}(\mathbf{x})x_i^2,
\end{eqnarray}
and insert it into the integral (\ref{00CS}).
By the $q$-analogue of the Stokes' formula (\ref{qStokes}), we obtain
\begin{align}\label{conCS}
0=&\frac{1}{Z^{rCS}} \int_{\mathbb{R}_+^N}d_q\mathbf{x} \left(\sum_{i=1}^{N}\frac{\partial}{\partial_qx_i}A_{t^{-1},i}(\mathbf{x})x_i^2\right)\prod_{i=1}^{N}
x_i^a e^{-\frac{\log^2x_i} {2\log q}} \Delta_{q,t}(\mathbf{x}) {_0\Phi_0^{(q,t)}(\mathbf{x,y})} \nonumber\\
=&\frac{1}{Z^{rCS}} \int_{\mathbb{R}_+^N}d_q\mathbf{x} \prod_{i=1}^{N} x_i^ae^{-\frac{\log^2x_i}{2\log q}} \Delta_{q,t}(\mathbf{x})\mathbf{Q}^{(rCS)}
(a;\mathbf{x}){_0\Phi_0^{(q,t)}(\mathbf{x,y})}\nonumber\\
=&\mathbf{L}^{(rCS)}(a;\mathbf{y})\langle{_0\Phi_0^{(q,t)}(\mathbf{x,y})}\rangle^{(rCS)},
\end{align}
where we have used (\ref{E200}) in the last line and
\begin{subequations}
\begin{align}
	\mathbf{Q}^{(rCS)}(a;\mathbf{x})=&\frac{t^{1-N}}{1-q}\left(p_1(\mathbf{x})-\sum_{i=1}^{N}A_{t,i}(\mathbf{x})q^{a+3/2}T_{q,i}(\mathbf{x})\right)\nonumber\\
	=&\frac{t^{1-N}}{1-q}p_1(\mathbf{x})+q^{a+3/2}t^{1-N}\left(\mathcal{E}_1(\mathbf{x})-\frac{1}{1-q}\{N\}_t\right),\\
	\mathbf{L}^{(rCS)}(a;\mathbf{y})=&t^{1-N}\left[W_{-1}^{(1)}(t^N;\mathbf{y})+q^{a+3/2}\left(\mathcal{E}_1(\mathbf{y})-\frac{1}{1-q}\{N\}_t\right)\right].\label{LCS}
\end{align}
\end{subequations}
From (\ref{L00cons}), we have
\begin{eqnarray}
	\mathbf{L}^{(rCS)}(a;\mathbf{y})=t^{1-N}\tilde{L}_0(-q^{a+3/2};\mathbf{y}),
\end{eqnarray}
which annihilates $\left({_0\phi}_0^{(q,t)}(-q^{a+3/2}\mathbf{y})\right)^{-1}$.

When taking the transformation $\mathbf{y}\rightarrow\mathbf{y}_n=(y_1,y_2,\cdots,y_n)$ with $1\le n\le N$, the constraint (\ref{conCS}) always holds.
It follows from Theorem \ref{UniqueL11} that
\begin{align}\label{00CS2}
	\langle{_0\Phi_0^{(q,t)}(\mathbf{x,y})}\rangle^{(rCS)}&=\left({_0\phi}_0^{(q,t)}(-q^{a+3/2}\mathbf{y})\right)^{-1}\nonumber\\
	&=\exp\left(\sum_{n=1}^{\infty}(-1)^{n-1}\frac{q^{(a+3/2)n}}{1-q^n}\frac{p_n(\mathbf{y})}{n}\right).
\end{align}
Expanding the both sides of (\ref{00CS2}) with $J_{\lambda}^{(q,t)}(\mathbf{y})$, we obtain
\begin{eqnarray}\label{00CSproof}
	&&\sum_{\lambda}\frac{J^{(q,t)}_{\lambda}(\mathbf{y})}{[t^{N}]_{\lambda}^{(q,t)}j_{\lambda}}\langle J^{(q,t)}_{\lambda}(\mathbf{x}) \rangle^{(rCS)}\nonumber\\
	&=&\sum_{\lambda} J^{(q,t)}_\lambda\left\{p_k=-\frac{(-q^{a+3/2})^k}{1-t^k}\right\}\frac{J^{(q,t)}_{\lambda}(\mathbf{y})}{j_{\lambda}}.
\end{eqnarray}
From (\ref{00CSproof}), we immediately confirm that the superintegrability relation (\ref{avCS}) holds. $\square$

We see that (\ref{00CS2}) provides a integral form for  $\left({_0\phi}_0^{(q,t)}(-q^{a+3/2}\mathbf{y})\right)^{-1}$.
We may give more general $q$-integrals
\begin{align}\label{ZsrCS}
\mathcal{Z}^{rCS}_{s,r}(\mathbf{a;b;y}):=&\frac{1}{Z^{rCS}} \int_{\mathbb{R}_+^N}d_q\mathbf{x} \prod_{i=1}^{N} x_i^ae^{-\frac{\log^2x_i}{2\log q}}
\Delta_{q,t}(\mathbf{x}){_s\Phi_r}^{(q,t)} (\mathbf{x,y})\nonumber\\
=&\hat{O}^{(s)}_{q,t}(\mathbf{a;y})\left(\hat{O}^{(r)}_{q,t}(\mathbf{b;y})\right)^{-1}\langle{_0\Phi_0^{(q,t)}(\mathbf{x,y})}\rangle^{(rCS)}\nonumber\\
=&\sum_{\lambda}\frac{\prod_{j=1}^{s}[a_j]^{(q,t)}_{\lambda}}{\prod_{k=1}^{r} [b_k]^{(q,t)}_{\lambda}}\frac{(-1)^{|\lambda|}q^{(a+3/2)
|\lambda|}q^{n(\lambda^T)}}{j_{\lambda}}J^{(q,t)}_\lambda\{\mathbf{y}\}.
\end{align}
Especially, for the case of $r=0$ in (\ref{ZsrCS}), it gives the $W$-representation
\begin{eqnarray}
	\mathcal{Z}^{rCS}_{s,0}(\mathbf{a;y})=\exp\left(\sum_{n=1}^{\infty}(-1)^{n-1}\frac{q^{(a+3/2)n}}{n}W_n^{(s)}(\mathbf{a;y})\right)\cdot 1.
\end{eqnarray}

\subsection{$q$-Selberg integral}
The Selberg integral is given by \cite{Selberg}
\begin{eqnarray}\label{Sel}
	Z^{S}=\int_{[0,1]^N}d\mathbf{x}\prod_{i=1}^{N}x_i^{a-1}(1-x_i)^{b-1}\Delta^{2\beta}(\mathbf{x}),
\end{eqnarray}
where $\operatorname{Re}(a),\operatorname{Re}(b)>0$.

We denote the Jack polynomials $J^{\beta}_{\lambda}=\lim_{t=q^{\beta}\to1}J^{(q,t)}_{\lambda}$.
The normalized averages of Jack polynomials are \cite{Kadell}
\begin{align}\label{avsel}
\langle J^{\beta}_{\lambda}(\mathbf{x})\rangle^{(S)} &=\frac{1}{Z^{S}} \int_{[0,1]^N}d\mathbf{x}\prod_{i=1}^{N}
x_i^{a-1}(1-x_i)^{b-1}\Delta^{2\beta}(\mathbf{x})J^{\beta}_{\lambda}(\mathbf{x})\nonumber\\
&=\frac{[N\beta]_{\lambda}^{\beta}[a+(N-1)\beta]_{\lambda}^{\beta}}{[a+b+2(N-1)\beta]_{\lambda}^{\beta}}J^{\beta}_{\lambda}\{p_k=\beta^{-1}\delta_{k,1}\},
\end{align}
where
\begin{align}
	[a]_{\lambda}^{\beta}&=\prod_{(i,j)\in\lambda}[a+j-1+(1-i)\beta]=\frac{J^{\beta}_{\lambda}\{p_k=\beta^{-1}a\}}{J^{\beta}_{\lambda}\{p_k=\beta^{-1}\delta_{k,1}\}}\nonumber\\
	&=\lim_{q\rightarrow 1}[q^a]^{(q,q^{\beta})}_{\lambda}/(1-q)^{|\lambda|}.
\end{align}

By taking the $q$-measure $d_q\mathbf{x}$ and the quantum deformations
\begin{eqnarray}
	\Delta^{2\beta}(\mathbf{x})&\rightarrow&\prod_{i=1}^{N} x_i^{(N-1)\log_q{t}} \Delta_{q,t}(\mathbf{x}),\nonumber\\ (1-x_i)^{b-1}&\rightarrow&(qx_i;q)_{b-1},
\end{eqnarray}
in the integral (\ref{Sel}),
one may obtain the $q$-Selberg integral \cite{Macdonald}
\begin{eqnarray}\label{qSel}
	Z^{qS}(a,b)=\int_{[0,1]^N}d_q\mathbf{x}w^{(qS)}(a,b;\mathbf{x}),
\end{eqnarray}
where the weight function is
\begin{eqnarray}
	w^{(qS)}(a,b;\mathbf{x})=\prod_{i=1}^{N} x_i^{a-1+(N-1)\log_q{t}} (qx_i;q)_{b-1}\Delta_{q,t}(\mathbf{x}).
\end{eqnarray}
We denote $Z^{qS}=Z^{qS}(a,b)$ and its normalized average is
\begin{eqnarray}
	\langle f(\mathbf{x})\rangle^{(qS)}= \frac{1}{Z^{qS}} \int_{[0,1]^N} d_q\mathbf{x}w^{(qS)}(a,b;\mathbf{x})f(\mathbf{x}).
\end{eqnarray}

Note that another $q$-analogue of the Selberg integral is \cite{Askey}
\begin{eqnarray}\label{aqSel}
	\tilde{Z}^{qS}=\int_{[0,1]^N}d_q\mathbf{x}\tilde{w}^{qS}(a,b;\mathbf{x})
\end{eqnarray}
with the weight function
\begin{eqnarray}
	\tilde{w}^{(qS)}(a,b;\mathbf{x})=\prod_{i=1}^{N}x_i^{a-1}(qx_i;q)_{b-1}\prod_{k=1-\beta}^{\beta}\prod_{1\le i<j\le N}(x_i-q^kx_j),
\end{eqnarray}
which is not the symmetric functions of $\mathbf{x}$.
The lemma of Kadell in Ref.\cite{Kadell} allows $\tilde{w}^{(qS)}(a,b;\mathbf{x})$ to be the symmetric form
\begin{align}
	&\prod_{i=1}^{N}x_i^{a-1}(qx_i;q)_{b-1}\prod_{1\le i<j\le N}\left((x_i-x_j) \prod_{k=1-\beta}^{\beta-1}(x_i-q^kx_j)\right)\nonumber\\
	=& (-1)^{N(N-1)/2}w^{(qS)}(a,b;\mathbf{x}) \Big|_{t=q^{\beta}}.
\end{align}
Thus the two $q$-integrals (\ref{qSel}) and (\ref{aqSel}) have the same normalized averages.

The superintegrability relation of the $q$-Selberg integral is \cite{Macdonald,Kadell}
\begin{align}\label{avSel}
	\langle J^{(q,t)}_{\lambda}(\mathbf{x}) \rangle^{(qS)}=&\frac{[t^N]^{(q,t)}_ {\lambda} [q^{a}t^{N-1}]^{(q,t)}_{\lambda}} {[q^{a+b}t^{2N-2}]
	^{(q,t)}_{\lambda}}J^{(q,t)}_{\lambda}\left\{p_k=\frac{1}{1-t^k}\right\}\nonumber\\
	=&\frac{J^{(q,t)}_\lambda\left\{p_k=\frac{1-t^{Nk}}{1-t^k}\right\}J^{(q,t)}_\lambda\left\{p_k=\frac{1-q^{ak}t^{(N-1)k}}{1-t^k}\right\}}
	{J^{(q,t)}_\lambda\left\{p_k=\frac{1-q^{(a+b)k}t^{2(N-1)k}}{1-t^k}\right\}},
\end{align}
where we have used (\ref{cij}) in the second line.

Let us give a new and concise proof of (\ref{avSel}).
First, we inserting total derivative operator
\begin{eqnarray}
	\sum_{i=1}^{N}\frac{\partial}{\partial_qx_i}A_{t^{-1},i}(\mathbf{x})(1-x_i)x_i
\end{eqnarray}
into the average
\begin{align}\label{00qS}
	\langle {_0\Phi_0}^{(q,t)}(\mathbf{x,y}) \rangle^{(qS)}&=\frac{1}{Z^{qS}} \int_{[0,1]^N}d_q\mathbf{x} w^{(qS)}(a,b;\mathbf{x}){_0\Phi_0}^{(q,t)}(\mathbf{x,y})\nonumber\\
	&=\sum_{\lambda}\frac{J^{(q,t)}_{\lambda}(\mathbf{y})}{[t^{N}]_{\lambda}^{(q,t)}j_{\lambda}}\langle J^{(q,t)}_{\lambda}(\mathbf{x}) \rangle^{(qS)}.
\end{align}
By the $q$-analogue of the Stokes' formula (\ref{qStokes}) and (\ref{E200}), we obtain
\begin{align}\label{conSel}
	0=&\frac{1}{Z^{qS}}\int_{[0,1]^N}d_q\mathbf{x}\left(\sum_{i=1}^{N}\frac{\partial}{\partial_qx_i}A_{t^{-1},i}(\mathbf{x})(1-x_i)x_i\right)w^{(qS)}(a,b;\mathbf{x}) {{_0\Phi_0}^{(q,t)}(\mathbf{x,y})} \nonumber\\
	=&\frac{1}{Z^{qS}}\int_{[0,1]^N} d_q\mathbf{x}w^{(qS)}(a,b;\mathbf{x})\mathbf{Q}^{(qS)}(a,b;\mathbf{x}) {{_0\Phi_0}^{(q,t)}(\mathbf{x,y})}\nonumber\\
	=&\mathbf{L}^{(qS)}(a,b;\mathbf{y})\langle{{_0\Phi_0}^{(q,t)}(\mathbf{x,y})}\rangle^{(qS)},
\end{align}
where
\begin{subequations}
\begin{align}
	&\mathbf{Q}^{(qS)}(a,b;\mathbf{x}) \nonumber \\
	=&\frac{t^{1-N}}{1-q}\left(-\sum_{i=1}^{N}A_{t,i}(\mathbf{x})(q^{a}-q^{a+b}x_i)t^{N-1}T_{q,i}(\mathbf{x})+\{N\}_t-p_1(\mathbf{x})\right)\nonumber\\
	=&t^{1-N}\left[q^{a}t^{N-1}(\mathcal{E}_1(\mathbf{x})-q^{b}\mathcal{E}_2(\mathbf{x}))+\frac{1-q^{a}t^{N-1}}{1-q}\{N\}_t+\frac{q^{a+b}t^{2N-2}-1}{1-q}p_1(\mathbf{x})\right],\\
	&\mathbf{L}^{(qS)}(a,b;\mathbf{y}) \nonumber\\ =&t^{1-N}\left[q^{a}t^{N-1}\mathcal{E}_1(\mathbf{y})+\frac{1-q^{a}t^{N-1}}{1-q}\{N\}_t
	-W^{(2)}_{-1}(t^{N},q^{a+b}t^{2N-2};\mathbf{y})\right].
\end{align}
\end{subequations}

From (\ref{LsLr}), we have
\begin{eqnarray}
	\mathbf{L}^{(qS)}(a,b;\mathbf{y})=-t^{1-N}\mathcal{L}_{(1,1)}(q^{a}t^{N-1};q^{a+b}t^{2N-2};1;\mathbf{y}),
\end{eqnarray}
which annihilates the hypergeometric function $_1\phi_1^{(q,t)}(q^{a}t^{N-1};q^{a+b}t^{2N-2};\mathbf{y})$.

When taking the transformation $\mathbf{y}\rightarrow\mathbf{y}_n=(y_1,y_2,\cdots,y_n)$ with $1\le n\le N$, the constraint (\ref{conSel}) always holds.
It follows from Theorem \ref{UniqueL11} that
\begin{eqnarray}\label{av00S}
	\langle{_0\Phi_0^{(q,t)}(\mathbf{x,y})}\rangle^{(qS)}={_1\phi_1^{(q,t)}}(q^{a}t^{N-1};q^{a+b}t^{2N-2};\mathbf{y}).
\end{eqnarray}
Expanding the both sides of (\ref{av00S}) with $J^{(q,t)}_{\lambda}(\mathbf{y})$, we immediately obtain the superintegrability relation (\ref{avSel}).

More generally, we may construct the partition functions
\begin{align}
	\mathcal{Z}^{qS}_{s,r}(\mathbf{a;b;y})&:=\frac{1}{Z^{qS}} \int_{[0,1]^N}d_q\mathbf{x} w^{(qS)}(a,b;\mathbf{x})
	{_s\Phi_r^{(q,t)}}(\mathbf{a;b;x,y})\nonumber\\
	&=\hat{O}^{(s)}_{q,t}(\mathbf{a;y})\left(\hat{O}^{(r)}_{q,t}(\mathbf{b;y})\right)^{-1}\langle{_0\Phi_0^{(q,t)}(\mathbf{x,y})}\rangle^{(qS)}\nonumber\\
	&={_{s+1}\phi_{r+1}^{(q,t)}}(q^at^{N-1},\mathbf{a};q^{a+b}t^{2N-2},\mathbf{b};\mathbf{y}),
\end{align}
where we have used (\ref{orep}) and (\ref{av00S}).

\subsection{$(q,t)$-deformed Hermite and Laguerre ensembles}

The $\beta$-deformed Hermite and Laguerre ensembles are given by \cite{Forrester08}
\begin{subequations}\label{ZHL}
\begin{eqnarray}
	Z^{H}&=&\int_{\mathbb{R}^N}d\mathbf{x} \prod_{i=1}^Ne^{-x_i^2}\Delta^{2\beta}(\mathbf{x}) ,\\
	Z^{L}&=&\int_{\mathbb{R}_+^N}d\mathbf{x}\prod_{i=1}^{N}x_i^{a-1} e^{-x_i}\Delta^{2\beta}(\mathbf{x}),\quad \operatorname{Re}(a)>0.
\end{eqnarray}
\end{subequations}
The superintegrability relations for $Z^H$ and $Z^L$ are \cite{Baker97,Fan}
\begin{subequations}\label{JackavHL}
\begin{eqnarray}
\langle J^{\beta}_{\lambda}(\mathbf{x})\rangle^{(H)} &=&\frac{1}{Z^H}\int_{\mathbb{R}^N}d\mathbf{x} \prod_{i=1}^Ne^{-x_i^2}
\Delta^{2\beta}(\mathbf{x})J^{\beta}_{\lambda}(\mathbf{x})\nonumber\\ &=&2^{-\lambda}[N\beta]^{(\beta)}_{\lambda}J^{\beta}_{\lambda}\{2\beta^{-1}\delta_{n,2}\}
=\frac{J^{\beta}_{\lambda}\{2\beta^{-1}\delta_{n,2}\}J^{\beta}_{\lambda}\{N\}}{J^{\beta}_{\lambda}\{2\beta^{-1}\delta_{n,1}\}},\\
\langle J^{\beta}_{\lambda}(\mathbf{x}) \rangle^{(L)}
&=&\frac{1}{Z^L}\int_{\mathbb{R}_+^N}d\mathbf{x}\prod_{i=1}^{N}x_i^{a-1} e^{-x_i} \Delta^{2\beta}(\mathbf{x})J^{\beta}_{\lambda}(\mathbf{x})\nonumber\\
&=&[a+1+(N-1)\beta]_{\lambda}^{(\beta)}[N\beta]^{(\beta)}_{\lambda}J^{\beta}_{\lambda}\{\beta^{-1}\delta_{n,1}\}\nonumber\\ &=&\frac{J^{\beta}_{\lambda}\{\beta^{-1}(a+1)+N-1\}J^{\beta}_{\lambda}\{N\}}{J^{\beta}_{\lambda}\{\beta^{-1}\delta_{n,1}\}}.
	\end{eqnarray}	
\end{subequations}

In order to discuss their $(q,t)$-deformed versions,
let us first recall the partition function
$\mathcal{N}=2$ supersymmetric theory with gauge group $U(N)$ on the 3-manifold $D^2\times_q S^1$ \cite{Cassia}
\begin{eqnarray}\label{ZDS}
Z_{D^2\times_q S^1}^{\mathcal{N}_f}= \oint_{\mathcal{C}}d\mathbf{x}\prod_{i=1}^N\prod_{k=1}^{\mathcal{N}_f}
(qu_kx_i;q)_{\infty}\prod_{i=1}^{N}x_i^{a-1+(N-1)\log_q{t}}\Delta_{q,t}(\mathbf{x}),
\end{eqnarray}
where $u_k$ for $k=1,\cdots,\mathcal{N}_f$ are the masses of the $\mathcal{N}_f$ fundamental anti-chiral fields, the contour $\mathcal{C}$ is product of $N$ copies
of the unit circle and $a\in\mathbb{C}$ which equals to $1$ when $\mathcal{N}_f=2$.

In addition, from the theory of multivariable Al-Salam and Carlitz polynomials, there is another $q$-analogue of Hermite ensemble which is called the $(q,t)$-deformed
Gaussian integral \cite{Baker00,Forrester25}
\begin{eqnarray}\label{ZqG}
	Z^{qG}=\int_{[c,1]^N}d_q\mathbf{x}w^{(qG)}(c;\mathbf{x})
\end{eqnarray}
with the weight function
\begin{eqnarray}
w^{(qG)}(c;\mathbf{x})=\prod_{1\le i<j\le N}\left((x_i-x_j) \prod_{k=1-\beta}^{\beta-1} (x_i-q^kx_j)\right)\prod_{i=1}^N
\frac{(qx_i;q)_{\infty}(qx_i/c;q)_{\infty}}{(q;q)_{\infty}(c;q)_{\infty}(q/c;q)_{\infty}}.
\end{eqnarray}
Its normalized average of the symmetric polynomial $f(\mathbf{x})$ is defined by
\begin{eqnarray}
	\langle f(\mathbf{x})\rangle^{(qG)}=\frac{1}{Z^{qG}}\int_{[c,1]^N}d_q\mathbf{x}w^{(qG)}(c;\mathbf{x})f(\mathbf{x}).
\end{eqnarray}

Inspired by the partition functions $Z_{D^2\times_q S^1}^{\mathcal{N}_f}$ (\ref{ZDS}) with $\mathcal{N}_f=1,2$ and $Z^{(qG)}$ (\ref{ZqG}),
we define $(q,t)$-analogues of Laguerre and Hermite ensembles by the $q$-integrals
\begin{subequations}\label{qZHZL}
\begin{align}
	Z^{qH}(u_1,u_2)&=\int_{[u_1^{-1},u_2^{-1}]^N}d_q\mathbf{x}w^{(qH)}(u_1,u_2;\mathbf{x}), \label{qZH}\\
	Z^{qL}(a,u)&=\int_{[0,u^{-1}]^N}d_q\mathbf{x}w^{(qL)}(a,u;\mathbf{x}), \label{qZL}
\end{align}
\end{subequations}
where the weight functions are
\begin{subequations}
\begin{eqnarray}
	w^{(qH)}(u_1,u_2;\mathbf{x})&=&\prod_{i=1}^{N} x_i^{(N-1)\log_q{t}}(qu_1x_i;q)_{\infty} (qu_2x_i;q)_{\infty}\Delta_{q,t}(\mathbf{x}) \nonumber\\
	&\propto& w^{(qG)}(u_1/u_2;u_1\mathbf{x}),\\
	w^{(qL)}(a,u;\mathbf{x})&=&\prod_{i=1}^{N} x_i^{a-1+(N-1)\log_q{t}} (qux_i;q)_{\infty}\Delta_{q,t}(\mathbf{x}) \nonumber\\
	&\propto&\lim_{b\rightarrow\infty}w^{(qS)}(a,b;u\mathbf{x}), \quad \operatorname{Re}a>0.
\end{eqnarray}
\end{subequations}

We denote $Z^{qH}=Z^{qH}(u_1,u_2)$ and $Z^{qL}=Z^{qH}(a,u)$.
The integral intervals in (\ref{qZHZL}) are selected to ensure that
\begin{subequations}
\begin{eqnarray}
	\langle f(\mathbf{x}) \rangle^{(qH)} &=&\frac{1}{Z^{qH}}\int_{[u_1^{-1},u_2^{-1}]^N}d_q\mathbf{x}w^{(qH)}(u_1,u_2;\mathbf{x})f(\mathbf{x})
	\nonumber\\
	&=&\langle f(u^{-1}\mathbf{x}) \rangle^{(qG)}\big|_{c=u_1/u_2},\\
	\langle f(\mathbf{x})\rangle^{(qL)} &=&\frac{1}{Z^{qL}}\int_{[0,u^{-1}]^N}d_q\mathbf{x}w^{(qL)}(a,u;\mathbf{x}) f(\mathbf{x})\nonumber\\
	&=&\lim_{b\rightarrow\infty}\langle f(u^{-1}\mathbf{x}) \rangle^{(qS)},\label{qSqL}
\end{eqnarray}
\end{subequations}
where we have used the property (\ref{qpro}).

The averages of Macdonald polynomials were conjectured in Ref.\cite{Cassia}
\begin{subequations}
\begin{eqnarray}
	\langle J_{\lambda}(\mathbf{x})\rangle^{(qH)} &=&[t^N]^{(q,t)}_{\lambda}J^{(q,t)}_\lambda\left\{p_k=\frac{u_1^{-k}+u_2^{-k}}{1-t^k}\right\}\nonumber\\
	&=&\frac{J^{(q,t)}_\lambda\left\{p_k=\frac{1-t^{Nk}}{1-t^k}\right\}}{J^{(q,t)}_\lambda\left\{p_k=\frac{1}{1-t^k}\right\}}
    J^{(q,t)}_\lambda\left\{p_k=\frac{u_1^{-k}+u_2^{-k}}{1-t^k}\right\},\label{avqH}\\
	\langle J_{\lambda}(\mathbf{x})\rangle^{(qL)} &=&\frac{J^{(q,t)}_\lambda \left\{p_k=\frac{1-t^{Nk}}{1-t^k}\right\}J^{(q,t)}_\lambda\left\{p_k
    =\frac{1-q^{ak}t^{(N-1)k}}{1-t^k}\right\}}{J^{(q,t)}_\lambda\left\{p_k=\frac{u^k}{1-t^k}\right\}}.\label{avqL}
	\end{eqnarray}
\end{subequations}
Note that (\ref{avqH}) was proved by the multivariable Al-Salam-Carlitz polynomials \cite{Forrester25}. (\ref{avqL}) can be checked easily by the limits (\ref{qSqL}) and (\ref{avSel}).

Let us give a new and concise proof of (\ref{avqH}).
We insert the total derivative operator
\begin{eqnarray}
	\sum_{i=1}^N\frac{\partial}{\partial_q x_i}A_{t^{-1},i}(\mathbf{x})(1-u_1x_i)(1-u_2x_i)
\end{eqnarray}
into the average $\langle _0\Phi_0^{(q,t)}(\mathbf{x,y})\rangle^{(qH)}$.
Then we have
\begin{align}\label{consqH}
	0=&\frac{1}{Z^{qH}} \int_{[u_1^{-1},u_2^{-1}]^N}d_q\mathbf{x} \left(\sum_{i}\frac{\partial}{\partial_q x_i}A_{t^{-1},i}(\mathbf{x})(1-u_1x_i)(1-u_2x_i) \right)\nonumber\\
	&\times w^{(qH)}(u_1,u_2;\mathbf{x}) {_0\Phi_0}^{(q,t)}(\mathbf{x,y})\nonumber\\
	=&\frac{1}{Z^{qH}} \int_{[u_1^{-1},u_2^{-1}]^N}d_q\mathbf{x} w^{(qH)}(u_1,u_2;\mathbf{x}) \mathbf{Q}^{(qH)}(u_1,u_2;\mathbf{x}){{_0\Phi_0}^{(q,t)}(\mathbf{x,y})}\nonumber\\
	=&\mathbf{L}^{(qH)}(u_1,u_2;\mathbf{y})\langle{{_0\Phi_0}^{(q,t)}(\mathbf{x,y})}\rangle^{(qH)},
\end{align}
where
\begin{subequations}
\begin{align}
	\mathbf{Q}^{(qH)}(u_1,u_2;\mathbf{x})=&\mathcal{E}_0(\mathbf{y})+u_1u_2\frac{t^{1-N}}{1-q}p_1(\mathbf{x})-(u_1+u_2)t^{1-N}\frac{\{N\}_t}{1-q},\\
	\mathbf{L}^{(qH)}(u_1,u_2;\mathbf{y})=&\frac{p_1(\mathbf{y})}{1-q}+u_1u_2t^{1-N}W_{-1}^{(1)} (t^N;\mathbf{y})-(u_1+u_2)t^{1-N}\frac{\{N\}_t}{1-q}.
\end{align}
\end{subequations}
Taking $\mathbf{a}=(0,0)$ and $\mathbf{w}=(u_1^{-1},u_2^{-1})$ in (\ref{L1010}), we have
\begin{eqnarray}
	\mathbf{L}^{(qH)}(u_1,u_2;\mathbf{y})=u_1u_2t^{1-N}\mathcal{L}_0(\mathbf{a,w};\mathbf{y}),
\end{eqnarray}
which annihilates ${_0\phi}_0^{(q,t)}(u_1^{-1}\mathbf{y}){_0\phi}_0^{(q,t)}(u_2^{-1}\mathbf{y})$.

When taking the transformation $\mathbf{y}\rightarrow\mathbf{y}_n=(y_1,y_2,\cdots,y_n)$ with $1\le n\le N$, the constraint (\ref{consqH}) always holds.
It follows from Theorem \ref{UniqueL11} that
\begin{eqnarray}\label{av00H}
	\langle{_0\Phi_0^{(q,t)}(\mathbf{x,y})}\rangle^{(qH)}&=&{_0\phi}_0^{(q,t)}(u_1^{-1}\mathbf{y}){_0\phi}_0^{(q,t)}(u_2^{-1}\mathbf{y})\nonumber\\
	&=&\exp\left(\sum_{n=1}^{\infty}\frac{u_1^{-n}+u_2^{-n}}{1-q^n}\frac{p_n(\mathbf{y})}{n}\right).
\end{eqnarray}
Expanding the both sides of (\ref{av00H}) with $J_{\lambda}^{(q,t)}(\mathbf{y})$, we obtain the superintegrability relation (\ref{avqH}).

More generally, we construct the partition functions
\begin{align}\label{ZqH}
	\mathcal{Z}^{qH}_{s,r}(\mathbf{a;b;y})=&\frac{1}{Z^{qH}} \int_{[u_1^{-1},u_2^{-1}]^N}d_q\mathbf{x} w^{(qH)}(a,b;\mathbf{x})
	{_s\Phi_r^{(q,t)}}(\mathbf{a;b;x,y})\nonumber\\
	=&\hat{O}^{(s)}_{q,t}(\mathbf{a;y})\left(\hat{O}^{(r)}_{q,t}(\mathbf{b;y})\right)^{-1}\langle{_0\Phi_0^{(q,t)}(\mathbf{x,y})}\rangle^{(qH)}\nonumber\\
	=&\sum_{\lambda}\frac{\prod_{j=1}^{s}[a_j]^{(q,t)}_{\lambda}}{\prod_{k=1}^{r} [b_k]^{(q,t)}_{\lambda}}J^{(q,t)}_\lambda\left\{p_k=\frac{u_1^{-k}+u_2^{-k}}
    {1-t^k}\right\}\frac{J^{(q,t)}_\lambda\{\mathbf{y}\}}{j_{\lambda}}.
\end{align}
Especially, for the case of $r=0$ in (\ref{ZqH}), it gives the $W$-representation
\begin{eqnarray}
	\mathcal{Z}^{qH}_{s,0}(\mathbf{a;y})=\exp\left(\sum_{n=1}^{\infty}\frac{u_1^{-n}+u_2^{-n}}{n}W_n^{(s)}(\mathbf{a;y})\right)\cdot1.
\end{eqnarray}

\section{A general $(q,t)$-deformed matrix model}
Let us construct the $(q,t)$-deformed matrix integral with a general weight function
\begin{eqnarray}\label{Zall}
	\tilde{Z}(a,c,u,v)=\int d_q\mathbf{x} \tilde{w}(a,c,u,v;\mathbf{x}),
\end{eqnarray}
where $u=(u_1,u_2)\in\mathbb{R}^2$, $v=(v_1,v_2)\in\mathbb{R}^2$, $\mathrm{Re}(a)\ge{-1}$ and
\begin{eqnarray}\label{Weall}
	\tilde{w}(a,c,u,v;\mathbf{x})=\prod_{i=1}^Nx_i^{a+(N-1)\log_q t}e^{-c\frac{(\log x_i)^2}{2\log q}}
\frac{(qu_1x_i;q)_{\infty} (qu_2x_i;q)_{\infty}}{(qv_1x_i;q)_{\infty} (qv_2x_i;q)_{\infty}}\Delta_{q,t}(\mathbf{x}).
\end{eqnarray}
Note that the integral domain is uncertain, but we may assume $\tilde{w}(a,c,u,v;\mathbf{x})$ (\ref{Weall}) always be zero at the boundary.

We define its normalized average for any symmetric polynomial $f(\mathbf{x})$
\begin{eqnarray}
	\langle f(\mathbf{x})\rangle= \frac{1}{\tilde{Z}(a,c,u,v)}\int d_q\mathbf{x}\tilde{w}(a,c,u,v;\mathbf{x})f(\mathbf{x}).
\end{eqnarray}

Let us insert the total derivative operator
\begin{eqnarray}
	\mathcal{T}_m(\mathbf{x})=\sum_{i=1}^N\frac{\partial}{\partial_q x_i}x_i^mA_{t^{-1},i}(\mathbf{x})(1-u_1x_i)(1-u_2x_i)
\end{eqnarray}
into $\langle{_0\Phi_0^{(q,t)}(\mathbf{x,y})}\rangle$, we obtain
\begin{eqnarray}\label{Consall}
	0&=&\int d_q\mathbf{x} \mathcal{T}_m(\mathbf{x})\tilde{w}(a,c,u,v;\mathbf{x}){_0\Phi_0^{(q,t)}}(\mathbf{x,y})\nonumber\\
	&=&\int d_q\mathbf{x} \tilde{w}(a,c,u,v;\mathbf{x}) \mathbf{Q}_m(a,c,u,v;\mathbf{x}){_0\Phi_0^{(q,t)}}(\mathbf{x,y})\nonumber\\
	&=&\mathbf{L}_m(a,c,u,v;\mathbf{y})\int d_q\mathbf{x} \tilde{w}(a,c,u,v;\mathbf{x}) {_0\Phi_0^{(q,t)}}(\mathbf{x,y}),
\end{eqnarray}
where
\begin{eqnarray}\label{Qall}
	&&\mathbf{Q}_m(a,c,u,v;\mathbf{x})\nonumber\\
	&=&\sum_{i=1}^Nx_i^{m-1}\frac{A_{t^{-1},i}(\mathbf{x})}{1-q}(1-u_1x_i)(1-u_2x_i)\nonumber\\
	&&-q^{m+a-c/2}(1-qv_1x_i)(1-qv_2x_i) x_i^{m-1-c}\frac{A_{t,i}(\mathbf{x})}{1-q}T_{q,i}(\mathbf{x})\nonumber\\
	&=&\bar{\mathcal{A}}_{m-1}(\mathbf{x})-(u_1+u_2)\bar{\mathcal{A}}_{m}(\mathbf{x})+u_1u_2\bar{\mathcal{A}}_{m+1}(\mathbf{x})\nonumber\\
	&&-q^{m+a-c/2}[\mathcal{A}_{m-1-c}(\mathbf{x})-q(v_1+v_2)\mathcal{A}_{m-c}(\mathbf{x})+q^2v_1v_2\mathcal{A}_{m+1-c}(\mathbf{x})]\nonumber\\
	&&+q^{m+a-c/2}[\mathcal{E}_{m-c}(\mathbf{x})-q(v_1+v_2)\mathcal{E}_{m-c+1}(\mathbf{x})+q^2v_1v_2\mathcal{E}_{m+2-c}(\mathbf{x})],
\end{eqnarray}
and the operators $\mathbf{L}_m(a,c,u,v)$ are to be determined. When taking the transformation $\mathbf{y}\rightarrow\mathbf{y}_n=(y_1,y_2,\cdots,y_n)$ with $1\le n\le N$,
the constraint (\ref{Consall})  always holds. It follows from Theorem \ref{UniqueL11} that the solution of (\ref{Consall}) is unique.

The operators $\mathbf{Q}_m(a,c,u,v;\mathbf{x})$ and $\mathbf{L}_m(a,c,u,v)$ are linearly composed of several homogeneous terms with degree $d_Q$ and $d_L$, respectively.
In order to achieve the hypergeometric constraints in section 2 from (\ref{Consall}), it requires that (i) $-1\le d_Q,d_L\le1$; (ii) $\bar{\mathcal{A}}_{m-1}(\mathbf{x})$
and $\mathcal{A}_{m-1}(\mathbf{x})$ with $m\le0$ should disappear in (\ref{Qall}). Therefore, it gives the following constraint conditions for the parameters:
\begin{align}\label{assume}
	-1\le m-1&\le 1,& -1\le m-c-1&\le 1,\nonumber\\
    (m+1)\cdot\theta\{u_1u_2=0\}&\le1,&
     m\cdot\theta\{u_1,u_2=0\}&\le1,\nonumber\\
    (m+1-c)\cdot\theta\{v_1v_2=0\}&\le1,&
    (m-c)\cdot\theta\{v_1,v_2=0\}&\le1,\nonumber\\
    (m-1)\cdot\theta\{c,m+a=0\}&\ge0,&  (m-c-1)\cdot\theta\{c,m+a=0\}&\ge0,
\end{align}
where $\theta\{P\}=0$ if $P$ is true and $\theta\{P\}=1$
if $P$ is false.
We enumerate all possible parameter values allowed by the constraint conditions (\ref{assume}) as follows.

(i) $u_2=v_2=c=0$ and $m=1$.

The $(q,t)$-deformed matrix integral (\ref{Zall}) becomes
\begin{eqnarray}\label{Z1}
	Z_{1}(a,u_1,v_1)=\int d_q\mathbf{x} \prod_{i=1}^Nx_i^{a+(N-1)\log_q t} \frac{(qu_1x_i;q)_{\infty} }{(qv_1x_i;q)_{\infty}}\Delta_{q,t}(\mathbf{x}).
\end{eqnarray}
From (\ref{Qall}), we have
\begin{subequations}
\begin{align}
	&\mathbf{Q}_1(a,0,u_1,v_1;\mathbf{x})\nonumber\\
	=&(t^{1-N}-q^{a+1})\frac{\{N\}_t}{1-q}+(q^{a+2}v_1t^{2N-2}-u_1)\frac{t^{1-N}}{1-q}p_1(\mathbf{x})+q^{a+1}\mathcal{E}_1(\mathbf{x})-q^{a+2}v_1\mathcal{E}_2(\mathbf{x})\nonumber\\
	=&(t^{1-N}-q^{a+1})\frac{\{N\}_t}{1-q}+q^{a+1}\mathcal{E}_1(\mathbf{x})-u_1t^{1-N}W_1^{(1)}\left(q^{a+2}t^{2N-2}v_1/u_1;\mathbf{x}\right).
\end{align}
Then the operator $\mathbf{L}_1(a,0,u_1,v_1;\mathbf{y})$ in (\ref{Consall}) becomes the hypergeometric constraint operator
\begin{align}
	&\mathbf{L}_1(a,0,u_1,v_1;\mathbf{y})\nonumber\\
	=&(t^{1-N}-q^{a+1})\frac{\{N\}_t}{1-q}+q^{a+1}\mathcal{E}_1(\mathbf{y})-u_1t^{1-N}W_{-1}^{(2)}\left(t^N;q^{a+2}t^{2N-2}v_1/u_1;\mathbf{y}\right)\nonumber\\
	=&-t^{1-N}\mathcal{L}_{(1,1)}(q^{a+1}t^{N-1};q^{a+2}t^{2N-2}\frac{v_1}{u_1};u_1^{-1};\mathbf{y}),
\end{align}
\end{subequations}
where we have used (\ref{L1r}). It follows from (\ref{L1rcons}) that
\begin{eqnarray}
	\langle{_0\Phi_0^{(q,t)}(\mathbf{x,y})}\rangle_1={_1\phi_1}(q^{a+1}t^{N-1};q^{a+2}t^{2N-2}\frac{v_1}{u_1};u_1^{-1}\mathbf{y}).
\end{eqnarray}
Thus, the superintegrability relation is
\begin{align}
	\langle J^{(q,t)}_{\lambda}\rangle_1
	=\frac{J^{(q,t)}_\lambda\left\{p_k=\frac{1-t^{Nk}}{1-t^k}\right\}J^{(q,t)}_\lambda\left\{p_k=\frac{1-(q^{a+1}t^{(N-1)})^k}{1-t^k}\right\}}
   {J^{(q,t)}_\lambda\left\{p_k=\frac{u_1^k-(q^{a+2}t^{2N-2}v_1)^k}{1-t^k}\right\}}.
\end{align}

(ii) $u_2=v_1=v_2=0$, $c=-1$ and $m=1$.

The $(q,t)$-deformed matrix integral (\ref{Zall}) becomes
\begin{eqnarray}\label{Z2}
	Z_{2}(a,u_1)=\int d_q\mathbf{x} \prod_{i=1}^Nx_i^{a+(N-1)\log_q t}e^{\frac{(\log x_i)^2}{2\log q}} (qu_1x_i;q)_{\infty} \Delta_{q,t}(\mathbf{x}).
\end{eqnarray}
From (\ref{Qall}), we have
\begin{subequations}
\begin{align}
	&\mathbf{Q}_1(a,-1,u_1,0;\mathbf{x})\nonumber\\
	=&t^{1-N}\frac{\{N\}_t}{1-q}-\left(q^{a+3/2}t^{2N-2}+u_1\right)\frac{t^{1-N}}{1-q}p_1(\mathbf{x})+q^{a+3/2}\mathcal{E}_2(\mathbf{x})\nonumber\\
	=&t^{1-N}\frac{\{N\}_t}{1-q}-u_1t^{1-N}W^{(1)}_1\left(-u_1^{-1}q^{a+3/2}t^{2N-2};\mathbf{x}\right).
\end{align}
Then the operator $\mathbf{L}_1(a,-1,u_1,0;\mathbf{y})$ in (\ref{Consall}) becomes the hypergeometric constraint operator
\begin{align}
	&\mathbf{L}_1(a,-1,u_1,0;\mathbf{y})\nonumber\\
	=&t^{1-N}\frac{\{N\}_t}{1-q}-u_1t^{1-N}W_{-1}^{(2)}\left(t^N,-u_1^{-1}q^{a+3/2}t^{2N-2};\mathbf{y}\right)\nonumber\\
	=&-u_1t^{1-N}\mathbf{Ad}^{-1}_{\hat{O}_{q,t}\left(-q^{a+3/2}t^{2N-2}u_1^{-1};\mathbf{y}\right)}L_0(0,u_1^{-1};\mathbf{y}).
\end{align}
\end{subequations}
It follows from (\ref{degrad}) that
\begin{align}
	\langle{_0\Phi_0^{(q,t)}(\mathbf{x,y})}\rangle_2&=\hat{O}^{-1}_{q,t}\left(-q^{a+3/2}t^{2N-2}u_1^{-1};\mathbf{y}\right){_0\phi_0}(u_1^{-1}\mathbf{y}),\nonumber\\
	&={_0\phi_1}(-q^{a+3/2}t^{2N-2}u_1^{-1};u_1^{-1}\mathbf{y}).
\end{align}
Thus, the superintegrability relation is
\begin{align}
	\langle J^{(q,t)}_{\lambda}\rangle_2
	&=\frac{J^{(q,t)}_\lambda\left\{p_k=\frac{1-t^{Nk}}{1-t^k}\right\}}{J^{(q,t)}_\lambda\left\{p_k=\frac{u_1^k-(-q^{a+3/2}t^{2N-2})^k}
    {1-t^k}\right\}}J^{(q,t)}_\lambda\left\{p_k=\frac{1}{1-t^k}\right\}.
\end{align}

(iii) $u_1=u_2=v_2=0$, $c=1$ and $m=2$.

The $(q,t)$-deformed matrix integral (\ref{Zall}) becomes
\begin{eqnarray}\label{Z3}
	Z_3(a,v_1)=\int d_q\mathbf{x} \prod_{i=1}^N x_i^{a+(N-1)\log_q t}e^{-\frac{(\log x_i)^2}{2\log q}} (qv_1x_i;q)^{-1}_{\infty} \Delta_{q,t}(\mathbf{x}).
\end{eqnarray}
From (\ref{Qall}), we have
\begin{subequations}
\begin{align}
	&\mathbf{Q}_2(a,1,0,v_1;\mathbf{x})\nonumber\\
	=&\left(1-q^{a+5/2}t^{2N-2}v_1\right)\frac{t^{1-N}}{1-q}p_1(\mathbf{x})-q^{a+3/2}\left(\frac{\{N\}_t}{1-q}-\mathcal{E}_1(\mathbf{x})\right)+q^{a+5/2}v_1\mathcal{E}_2(\mathbf{x})\nonumber\\
	=&-q^{a+3/2}\left(\frac{\{N\}_t}{1-q}-\mathcal{E}_1(\mathbf{x})\right)-t^{1-N}W_{1}^{(1)}\left(-q^{a+5/2}t^{2N-2}v_1;\mathbf{x}\right).
\end{align}
Then the operator $\mathbf{L}_2(a,1,0,v;\mathbf{y})$ in (\ref{Consall}) becomes the hypergeometric constraint operator
\begin{align}
	&\mathbf{L}_2(a,1,0,v;\mathbf{y})\nonumber\\
	=&-q^{a+3/2}\left(\frac{\{N\}_t}{1-q}-\mathcal{E}_1(\mathbf{y})\right)-t^{1-N}W_{-1}^{(2)}\left(t^N,-q^{a+5/2}t^{2N-2}v_1;\mathbf{y}\right)\nonumber\\
	=&-t^{1-N}\mathbf{Ad}^{-1}_{\hat{O}_{q,t}\left(-q^{a+5/2}t^{2N-2}v_1;\mathbf{y}\right)}\tilde{L}_0(q^{a+3/2}t^{N-1};\mathbf{y}).
\end{align}
\end{subequations}
It follows from (\ref{degrad}) that
\begin{eqnarray}
	\langle{_0\Phi_0^{(q,t)}(\mathbf{x,y})}\rangle_3=\hat{O}_{q,t}^{-1}\left(-q^{a+5/2}t^{2N-2}v_1;\mathbf{y}\right)\left({_0\phi_0}(q^{a+3/2}t^{N-1}\mathbf{y})\right)^{-1}.
\end{eqnarray}
Thus, the superintegrability relation is
\begin{eqnarray}
\langle J^{(q,t)}_{\lambda}\rangle_3 =\frac{J^{(q,t)}_\lambda\left\{p_k=-\frac{(q^{a+3/2}t^{N-1})^k}{1-t^k}\right\}}{J^{(q,t)}_\lambda\left\{p_k=\frac{1-(-q^{a+5/2}t^{2N-2})^kv^k_1}{1-t^k}\right\}}
J^{(q,t)}_\lambda\left\{p_k=\frac{1-t^{Nk}}{1-t^k}\right\}.
\end{eqnarray}

(iv) $a=c=m=0$.

The $(q,t)$-deformed matrix integral (\ref{Zall}) becomes
\begin{eqnarray}\label{Z4}
	Z_{4}(u,v)=\int d_q\mathbf{x} \prod_{i=1}^Nx_i^{(N-1)\log_q t} \frac{(qu_1x_i;q)_{\infty}(qu_2x_i;q)_{\infty}}{(qv_1x_i;q)_{\infty}(qv_2x_i;q)_{\infty}} \Delta_{q,t}(\mathbf{x}).
\end{eqnarray}
From (\ref{Qall}), we have
\begin{align}
	&\mathbf{Q}_0(0,0,u,v;\mathbf{x})\nonumber\\
	=&[q(v_1+v_2)-(u_1+u_2)t^{1-N}]\frac{\{N\}_{t}}{1-q}+\left(u_1u_2-q^2t^{2N-2}v_1v_2\right)\frac{t^{1-N}}{1-q}p_1(\mathbf{x})\nonumber\\
	&+\mathcal{E}_0(\mathbf{x})-q(v_1+v_2)\mathcal{E}_1(\mathbf{x})+q^2v_1v_2\mathcal{E}_2(\mathbf{x})\nonumber\\
	=&-q(v_1+v_2)\mathcal{E}_{1}(\mathbf{x})-\left[u_1+u_2-qt^{N-1}(v_1+v_2)\right]\frac{t^{1-N}}{1-q}\{N\}_{t}\nonumber\\
	&+u_1u_2t^{1-N}W_{1}^{(1)}\left(q^2t^{2N-2}\frac{v_1v_2}{u_1u_2};\mathbf{x}\right)+W_{-1}^{(1)}(t^{N};\mathbf{x}).
\end{align}
Then the operator $\mathbf{L}_0(0,0,u,v;\mathbf{y})$ in (\ref{Consall}) becomes the hypergeometric constraint operator
\begin{align}
	&\mathbf{L}_0(0,0,u,v;\mathbf{y})\nonumber\\
	=&-q(v_1+v_2)\mathcal{E}_1(\mathbf{y})-\left[u_1+u_2-qt^{N-1}(v_1+v_2)\right]\frac{t^{1-N}}{1-q}\{N\}_{t}\nonumber\\
	&+u_1u_2t^{1-N}W_{-1}^{(2)}\left(t^N,q^2t^{2N-2}\frac{v_1v_2}{u_1u_2};\mathbf{y}\right)+\frac{1}{1-q}p_1(\mathbf{y})\nonumber\\
	=&u_1u_2t^{1-N}\mathbf{Ad}^{-1}_{\hat{O}_{q,t}\left(q^2t^{2N-2}\frac{v_1v_2}{u_1u_2};\mathbf{y}\right)}\mathcal{L}_0(\mathbf{a,w;y}),
\end{align}
where $\mathbf{w}=(u_1^{-1},u_2^{-1})$ and $\mathbf{a}=(qt^{N-1}v_1/u_2,qt^{N-1}v_2/u_1)$.

It follows from (\ref{L1010}) that
\begin{align}
	\langle{_0\Phi_0^{(q,t)}(\mathbf{x,y})}\rangle_4=&\hat{O}_{q,t}^{-1}\left(q^2t^{2N-2}\frac{v_1v_2}{u_1u_2};\mathbf{y}\right)\nonumber\\
	&\cdot\left[{_1\phi_0}\left(qt^{N-1}\frac{v_1}{u_2};u_1^{-1}
\mathbf{y}\right){_1\phi_0}\left(qt^{N-1}\frac{v_2}{u_1};u_2^{-1}\mathbf{y}\right)\right].
\end{align}
Thus, the superintegrability relation is
\begin{eqnarray}
	\langle J^{(q,t)}_{\lambda}\rangle_4
	 =\frac{J^{(q,t)}_\lambda\left\{p_k=\frac{(u_1^k+u_2^k)-(v_1^k+v_2^k)q^kt^{(N-1)k}}{1-t^k}\right\}}{J^{(q,t)}_\lambda
     \left\{p_k=\frac{u_1^ku_2^k-(q^2t^{2N-2})^kv^k_1v^k_2}{1-t^k}\right\}}J^{(q,t)}_\lambda\left\{p_k=\frac{1-t^{Nk}}{1-t^k}\right\}.
\end{eqnarray}

We see that the above cases (i)-(iv) coincide with those (see \textbf{SI}.7-\textbf{SI}.10) in \cite{CassiaNew}, and other cases in \cite{CassiaNew}
can be obtained by certain parameter degradation from cases (i)-(iv).

Especially, the $(q,t)$-deformed matrix models in section 3 are
\begin{subequations}
\begin{align}
	Z^{rCS}(a)&=\lim_{v_1\to0}Z_3(a-(N-1)\log_qt,v_1),\\Z^{qS}(a;b)&=\lim_{u_1\to0 }Z_1(a-1,u_1,u_1q^{b-1}),\\Z^{qH}(u_1,u_2)&=\lim_{v_1,v_2\to0}Z_4(u,v).
\end{align}
\end{subequations}

\section{Conclusions}
We have investigated the $(q,t)$-deformed hypergeometric functions (\ref{hyper}) and presented their representations (\ref{orep}) and (\ref{wrep}) associated
with the $O$-operator (\ref{Oop}) and $W$-operators (\ref{wowo}). By the $W$-operators, we have constructed the constraints (\ref{ConsXY})
for ${_s\Phi_r^{(q,t)}(\mathbf{a;b;x;y})}$, (\ref{ConsW+-0}) for ${_s\phi_r^{(q,t)}(\mathbf{a;b;x})}$,  and (\ref{L1010})
for ${_1\phi_0^{(q,t)}(a_1;w_1\mathbf{x})}{_1\phi_0^{(q,t)}(a_2;w_2\mathbf{x})}$. We have proved the uniqueness to the solutions of the hypergeometric
constraints (\ref{ConsW+-0}) and (\ref{L1010}) with $k=0$.

We have proposed a concise method to prove the superintegrability relations for some $(q,t)$-deformed matrix models as follows.
First, we identify a specific $(q,t)$-deformed integral and its normalized average of ${_0\Phi_0^{(q,t)}(\mathbf{x,y})}$.
Then, by using the $q$-analogue of the Stokes' formula (\ref{qStokes}), we give a single constraint of $\langle {_0\Phi_0^{(q,t)}(\mathbf{x,y})}\rangle$.
Due to the uniqueness to the solutions of the hypergeometric constraints, we obtain that
$\langle {_0\Phi_0^{(q,t)}(\mathbf{x,y})}\rangle$ is identical to certain $(q,t)$-deformed hypergeometric function.
Finally, by expanding the average $\langle {_0\Phi_0^{(q,t)}(\mathbf{x,y})} \rangle$, we obtain the superintegrability relation for the given matrix integral.

We focused on the refined unknot Chern-Simons matrix model (\ref{rCS}), $q$-Selberg integral (\ref{qSel}) and $(q,t)$-deformed Hermite and Laguerre ensembles (\ref{qZHZL}).
Their superintegrability relations (\ref{avCS}), (\ref{avSel}) and (\ref{avqH}) can be easily proved by our method.
In addition, inspired by above three $(q,t)$-deformed  matrix models, we have proposed a general $(q,t)$-deformed integral (\ref{Zall}).
We gave the constraint conditions (\ref{assume}) for the parameters such that the constraints (\ref{Consall}) reduce to the hypergeometric constraints.
By (\ref{assume}), four $(q,t)$-deformed integrals were obtained naturally from (\ref{Zall}).
These integrals coincide with the cases in Ref.\cite{CassiaNew}, and their superintegrability relations can be easily derived from our hypergeometric constraints.

\section*{Acknowledgments}
This work is supported by the National Natural Science Foundation of China (Nos. 12375004 and 12205368) and the
Fundamental Research Funds for the Central Universities, China (No. 2024ZKPYLX01).

\appendix

\section{$q$-derivative and $q$-integral}
Let us start from the $q$-derivative $\frac{d}{d_q x}$ with $q\in(0,1)$ which is defined by \cite{Exton}
\begin{eqnarray}\label{deri}
	\frac{d}{d_q x}(f(x))=\frac{f(x)-f(qx)}{(1-q)x}=\frac{1-q^{x\partial_{x}}}{(1-q)x}(f(x)).
\end{eqnarray}
The $q$-integral for the function $f(x)$ is given by \cite{Jackson}
\begin{eqnarray}\label{int}
	\int_0^u f(x)d_q x=(1-q)\sum_{k=0}^{\infty}uq^kf(q^ku),
\end{eqnarray}
which is defined as the inverse operation of the $q$-derivative $\frac{d}{d_q x}$, i.e.,
\begin{eqnarray}
	\int_0^u \frac{d}{d_q x}(f(x))d_q x=f(u)-f(0).
\end{eqnarray}
The definitions (\ref{deri}) and (\ref{int}) can be lifted to the multivariable case \cite{Askey}
\begin{align}
	\frac{\partial}{\partial_q x_i}(f(\mathbf{x}))=&\frac{1-q^{x_i\partial_{x_i}}}{(1-q)x_i}(f(\mathbf{x})) ,\nonumber\\
	\int_{[0,u]^N}d_q\mathbf{x}f(\mathbf{x})=&(1-q)^Nu^N\sum_{\alpha_j\in\mathbb{Z}_{\ge 0},j\in I_N} q^{\sum_{j=1}^N\alpha_j} f(q^{\alpha_1}u,\dots,q^{\alpha_N}u),\label{qmeasure}
\end{align}
where $1\le i\le N$, $I_N=\{1,2,\cdots,N\}$ and $f(\mathbf{x})$ is defined on the cube $\mathbf{x}=(x_1,\cdots,x_N)\in[0,u]^N$ with $u\in\mathbb{R}_+$.
Then we call $d_q\mathbf{x}$ $q$-measure in this paper.

From the definition of the $q$-measure (\ref{qmeasure}), it is easy to check that the $q$-integral satisfies the following properties
\begin{eqnarray}\label{qpro}
	\int_{[0,u]^N}d_q\mathbf{x}f(\mathbf{x})=u^{N}\int_{[0,1]^N}d_q\mathbf{x}f(u\mathbf{x}),
\end{eqnarray}
and
\begin{align}\label{qStokes}
\int_{[0,u]^N}d_q\mathbf{x}\frac{\partial}{\partial_q x_i}(f(\mathbf{x}))=& (1-q)^{N-1}u^{N-1}\sum_{\alpha_j\in\mathbb{Z}_{\ge 0},j\in I_N\setminus\{i\}}
q^{\sum_{j\in I_N\setminus\{i\}}\alpha_j} \nonumber\\
&\times f(q^{\alpha_1}u,\dots,q^{\alpha_{i-1}} u,x_{i}u,q^{\alpha_{i+1}}u,\cdots,q^{\alpha_N}u)|^{1}_{x_i=0}.
\end{align}
It is clear that when taking proper $f(\mathbf{x})$ such that $f(\mathbf{x})=0$ at the hyperplanes $x_i=0$ and $x_i=u$, the integral on the left side of (\ref{qStokes}) will be zero.
Thus we call (\ref{qStokes}) the $q$-analogue of the Stokes' formula in this paper.

}	
\end{document}